\documentclass[%
 reprint,
 superscriptaddress,
 amsmath,
 amssymb,
 aps,
 floatfix,
]{revtex4-2}

\usepackage{graphicx}
\usepackage{dcolumn}
\usepackage{bm}
\usepackage{hyperref}

\def\be{\begin{equation}}
\def\ee{\end{equation}}
\def\bea{\begin{eqnarray}}
\def\eea{\end{eqnarray}}

\usepackage{siunitx}
\usepackage{color}
\usepackage{comment}

\definecolor{navyblue}{rgb}{0,0,0.5} 
\definecolor{forestgreen}{rgb}{0,0.6,0}

\newcommand{\av}[1]{\left\langle#1\right\rangle}

\usepackage[dvipsname]{xcolor}
\usepackage{ulem}

\hypersetup{
    colorlinks=false,
    linkcolor=yellow,
    filecolor=magenta,      
    urlcolor=cyan,
}

\begin{document}

\preprint{APS/123-QED}

\title{
Observation of Universal Hall Response in Strongly Interacting Fermions
}

\affiliation{Department of Physics and Astronomy, University of Florence, 50019 Sesto Fiorentino, Italy}
\affiliation{Istituto Nazionale di Ottica del Consiglio Nazionale delle Ricerche (CNR-INO), Sezione di Sesto Fiorentino, 50019 Sesto Fiorentino, Italy}
\affiliation{European Laboratory for Non-Linear Spectroscopy (LENS), 50019 Sesto Fiorentino, Italy}
\affiliation{Université Grenoble Alpes, CNRS, LPMMC, 38000 Grenoble, France}
\affiliation{Department of Quantum Matter Physics, University of Geneva, 1211 Geneva, Switzerland}
\affiliation{Department of Engineering, Campus Bio-Medico University of Rome, 00128 Rome, Italy}
\affiliation{Université Grenoble Alpes, CEA, IRIG-MEM-L\_SIM, 38000 Grenoble, France}

\author{T.-W. Zhou}
\affiliation{Department of Physics and Astronomy, University of Florence, 50019 Sesto Fiorentino, Italy}
\affiliation{Istituto Nazionale di Ottica del Consiglio Nazionale delle Ricerche (CNR-INO), Sezione di Sesto Fiorentino, 50019 Sesto Fiorentino, Italy}

\author{G. Cappellini}
\affiliation{Istituto Nazionale di Ottica del Consiglio Nazionale delle Ricerche (CNR-INO), Sezione di Sesto Fiorentino, 50019 Sesto Fiorentino, Italy}
\affiliation{European Laboratory for Non-Linear Spectroscopy (LENS), 50019 Sesto Fiorentino, Italy}

\author{D. Tusi}
\affiliation{European Laboratory for Non-Linear Spectroscopy (LENS), 50019 Sesto Fiorentino, Italy}

\author{L. Franchi}
\affiliation{Department of Physics and Astronomy, University of Florence, 50019 Sesto Fiorentino, Italy}

\author{J. Parravicini}
\affiliation{Department of Physics and Astronomy, University of Florence, 50019 Sesto Fiorentino, Italy}
\affiliation{Istituto Nazionale di Ottica del Consiglio Nazionale delle Ricerche (CNR-INO), Sezione di Sesto Fiorentino, 50019 Sesto Fiorentino, Italy}
\affiliation{European Laboratory for Non-Linear Spectroscopy (LENS), 50019 Sesto Fiorentino, Italy}

\author{C. Repellin}
\affiliation{Université Grenoble Alpes, CNRS, LPMMC, 38000 Grenoble, France}

\author{S. Greschner}
\affiliation{Department of Quantum Matter Physics, University of Geneva, 1211 Geneva, Switzerland}

\author{M. Inguscio}
\affiliation{Department of Engineering, Campus Bio-Medico University of Rome, 00128 Rome, Italy}
\affiliation{Istituto Nazionale di Ottica del Consiglio Nazionale delle Ricerche (CNR-INO), Sezione di Sesto Fiorentino, 50019 Sesto Fiorentino, Italy}
\affiliation{European Laboratory for Non-Linear Spectroscopy (LENS), 50019 Sesto Fiorentino, Italy}

\author{T. Giamarchi}
\affiliation{Department of Quantum Matter Physics, University of Geneva, 1211 Geneva, Switzerland}

\author{M. Filippone}
\affiliation{Université Grenoble Alpes, CEA, IRIG-MEM-L\_SIM, 38000 Grenoble, France}

\author{J. Catani}
\affiliation{Istituto Nazionale di Ottica del Consiglio Nazionale delle Ricerche (CNR-INO), Sezione di Sesto Fiorentino, 50019 Sesto Fiorentino, Italy}
\affiliation{European Laboratory for Non-Linear Spectroscopy (LENS), 50019 Sesto Fiorentino, Italy}

\author{L. Fallani}
\email{fallani@lens.unifi.it}
\affiliation{Department of Physics and Astronomy, University of Florence, 50019 Sesto Fiorentino, Italy}
\affiliation{Istituto Nazionale di Ottica del Consiglio Nazionale delle Ricerche (CNR-INO), Sezione di Sesto Fiorentino, 50019 Sesto Fiorentino, Italy}
\affiliation{European Laboratory for Non-Linear Spectroscopy (LENS), 50019 Sesto Fiorentino, Italy}


\begin{abstract}
The Hall effect, which originates from the motion of charged particles in magnetic fields, has deep consequences for the description of materials, extending far beyond condensed matter. Understanding such an effect in interacting systems represents a fundamental challenge, even for small magnetic fields. In this work, we used an atomic quantum simulator in which we tracked the motion of ultracold fermions in two-leg ribbons threaded by artificial magnetic fields. Through controllable quench dynamics, we measured the Hall response for a range of synthetic tunneling and atomic interaction strengths. We unveil a universal interaction-independent behavior above an interaction threshold, in agreement with theoretical analyses. The ability to reach hard-to-compute regimes demonstrates the power of quantum simulation to describe strongly correlated topological states of matter.
\end{abstract}

\maketitle


Since its first observation in 1879~\cite{Edwin}, the Hall effect has been an extraordinary tool for understanding solid-state systems~\cite{popovic_hall_2003}. This phenomenon is a macroscopic manifestation of the motion of charge carriers in materials subjected to a magnetic field $B$, generating an electric field $E_y$ perpendicular to the longitudinal current $J_x$ flowing in the system. At a small magnetic field, the Hall coefficient $R_{\rm H}=E_y/(B J_x)$ permits the extraction of the effective charge $q$ and carrier density $n$, because $R_{\rm H}\simeq -1/nq$ in conventional conductors. The Hall effect has widespread applications in metrology and materials science, such as sensitive measurements of magnetic fields and resistance standards based on its quantized behavior at large $B$~\cite{PhysRevLett.45.494}. The modern understanding of the Hall effect establishes it as a manifestation of robust geometric properties of quantum systems: Fermi-surface curvature of metals under weak magnetic fields~\cite{ong91_geometric_hall,tsuji58_hall_effect_cubic}, Berry curvature of anomalous Hall systems~\cite{xiao2010_berry_phase_review}, and topological invariants of band insulators~\cite{thouless1982_quantized_hall_conductance}. Studies of Hall responses are ubiquitous in fields that address topological quantum matter~\cite{RevModPhys.82.3045} and synthetic realizations thereof~\cite{goldman2016topological,RevModPhys.91.015006}.

\begin{figure}[t!]
	\centering
	\includegraphics[width=\columnwidth]{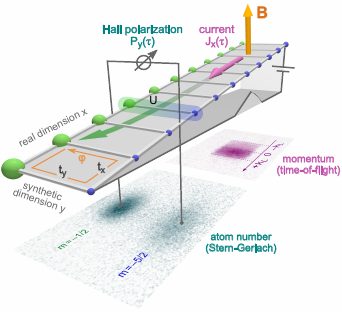}
	\caption{\textbf{Experimental scheme.} A synthetic ladder is realized by trapping fermionic $^{173}$Yb atoms in a 1D optical lattice and coupling their nuclear spins $m_{\rm F}=-1/2$ and $m_{\rm F}=-5/2$ via two-photon Raman transitions. The position-dependent phase of the Raman coupling simulates a magnetic field $B$ with Aharonov-Bohm phase $\varphi$ per unit cell. An atomic current is activated by tilting the ladder with an optical gradient, equivalent to a constant electric field $E_x$.~The radius difference of the green and blue spheres illustrates the leg population imbalance (Hall polarization) induced by the Hall drift.~The time-dependent longitudinal current $J_x(\tau)$ and the Hall polarization $P_y(\tau)$ are measured with time-of-flight imaging and optical Stern-Gerlach detection, respectively (typical acquisitions are shown below the ladder).}
	\label{fig1}
\end{figure}

However, when interactions are present among the carriers, understanding the Hall coefficient becomes a theoretical challenge. At large magnetic fields, interactions lead to the fractional quantum Hall effect~\cite{tsui_FQHE}, where the quantization of $R_{\rm H}B$ to fractions of $h/e^2$ (where $h$ is Planck's constant) reveals the emergence of elementary excitations with fractional charge and anyonic statistics~\cite{laughlin1983,Yoshioka}. For small fields, the connection of $R_{\rm H}$ with carrier densities and topological invariants is lost, leading to numerous theoretical attempts~\cite{RevModPhys.91.011002,PhysRevB.4.1566,PhysRevB.75.195123,PhysRevLett.121.066601,PhysRevLett.85.377,PhysRevB.55.3907,PhysRevLett.70.2004,PhysRevB.91.024507} to understand the effects of many-body correlations on this quantity. On the experimental side this complexifies the interpretation of anomalous temperature dependence and sign changes of $R_{\rm H}$ in the normal phase of cuprates~\cite{hagen_anomalous_1990,badoux_change_2016}, disordered superconducting films~\cite{smith_sign_1994} and organic compounds~\cite{PhysRevLett.84.2670,PhysRevLett.84.2674}. Numerical progress has recently allowed a reliable calculation of the Hall coefficient~\cite{PhysRevLett.122.083402} in a quasi-one-dimensional (1D) geometry and predicted a threshold of interactions above which the Hall coefficient becomes interaction-independent and thus universal.

In this context, ultracold atoms in optical lattices provide an opportunity to gain insight into the fundamental aspects of interacting Hall systems, owing to their flexibility and controllability. A notable recent advance was the realization of artificial magnetic fields in optical lattices, through various schemes including laser-induced tunneling, Floquet engineering and synthetic dimensions~\cite{RevModPhys.83.1523,Goldman2014,goldman2016topological, RevModPhys.91.015005, Ozawa2019}. Until now, these schemes have been exploited to explore single-particle~\cite{Mancini1510,Aidelsburger2015,genkina2019imaging,chalopin_probing_2020} and few-body~\cite{Greiner2017,Greiner2023} phenomena, while the observation of strongly correlated many-body effects triggered by interactions has remained elusive.

In this work, we report on the measurement of the Hall response in a quantum simulator with strongly interacting ultracold fermions. By controlling the repulsion between particles, we obtained experimental evidence of the universal response that is predicted at a large interaction strength. We used a synthetic dimension to engineer a two-leg ladder whose plaquettes are threaded by a synthetic magnetic flux $\varphi$ (see Fig.~\ref{fig1}). We monitored the real-time dynamics of the system after the instantaneous quench of a linear potential, which tilts the lattice along $\hat{x}$ and mimics the action of a longitudinal electric field $E_x$. We observe that the combined action of $E_x$ and $\varphi$ triggers a longitudinal current $J_x$, accompanied by the Hall polarization of the system along the transverse direction $P_y$. Even though the dynamics of $J_x$ and $P_y$ strongly depend on microscopic ladder parameters, we observe that a proxy of the Hall coefficient, the Hall imbalance~\cite{PhysRevLett.122.083402}
\begin{equation}\label{eq1}
	\Delta_{\rm H}=\left|\frac{P_y}{J_x}\right|
\end{equation}
converges towards an interaction-independent value for large atomic repulsions. Our observations quantitatively agree with theoretical calculations and confirm the predictions reported in~\cite{PhysRevLett.122.083402}. Our results showcase the importance of interactions in Hall systems, paving the way towards the investigation of strongly correlated effects in topological phases of synthetic quantum matter.

\subsection*{Making and probing a synthetic ladder}

Our experiment exploits an ultracold Fermi gas of $^{173}$Yb atoms initially polarized in the $\lvert F=5/2,\,m_{\rm F}=-5/2\rangle$ hyperfine state. The atoms are trapped in a 1D optical lattice, which allows real tunneling between different sites along direction $\hat{x}$. An additional 2D lattice (not shown in Fig.~\ref{fig1}) freezes the atomic motion along the orthogonal real-space directions, forming a 2D array of fermionic tubes. By adiabatically activating the coherent Raman couplings between nuclear spin states $\lvert m_{\rm F}=-1/2\rangle$ and $\lvert m_{\rm F}=-5/2\rangle$ (denoted $m=1,2$ respectively), our system realizes a two-leg ladder~\cite{Mancini1510}, in which the nuclear spins act as different sites along a synthetic dimension $\hat{y}$ (see Fig.~\ref{fig1}). The system is described by a two-leg version of the interacting Harper-Hofstadter Hamiltonian
\begin{align}\label{eq2}
	H=&-t_x\sum_{j,m}\left[a^{\dagger}_{j,m}a_{j+1,m}+\textrm{h.c.}\right]\nonumber\\
	&-t_y\sum_{j}\left[e^{i\varphi j}a^\dagger_{j,1}a_{j,2}+\textrm{h.c.}\right]+U\sum_{j}n_{j,1}n_{j,2}
\end{align}
where $a_{j,m}$ ($a^{\dagger}_{j,m}$) is the fermionic annihilation (creation) operator on site $(j,m)$ in the real and synthetic ($m=1,\,2$) dimensions, $n_{j,m}=a^\dagger_{j,m}a_{j,m}$, and h.c. is the Hermitian conjugate operator. Here, $t_x$ is the nearest-neighbor tunneling amplitude and $U>0$ is the ``on-rung'' interaction energy between two atoms with different nuclear spin in the same real-lattice site. The coupling between two spin states $t_y e^{i\varphi j}$ is interpreted as a tunneling along the synthetic dimension, whereby the position-dependent phase simulates the effect of a static magnetic flux $\varphi$ threading the ladder; in our experiments, $|\varphi|=0.32\pi$. A residual harmonic confining potential results in an additional term $H_{\rm conf.}=V_x\sum_{j,m}j^2n_{j,m}$, with the confinement strength $V_x=0.01t_x$. The atomic repulsion $U$ is controlled, independently from $t_x$ and $t_y$, by changing the radial confinement of fermionic tubes via the 2D lattice depth; to keep $V_x$ constant while changing $U$, we added a weak double-well potential along direction $\hat{x}$, compensating the trapping frequency by adjusting the potential barrier between the two wells~\cite{SS}.

\begin{figure}[t!]
	\centering
	\includegraphics[width=\columnwidth]{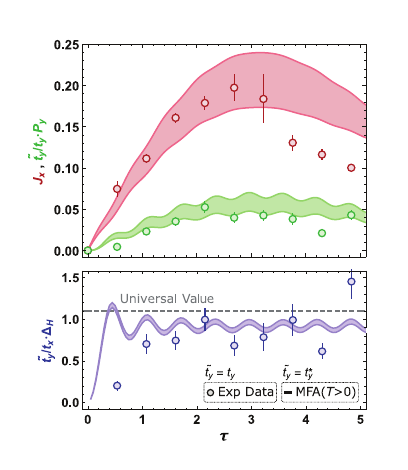}
	\caption{\textbf{Time evolution of the particle current $J_x$, Hall polarization $P_y$, and Hall imbalance $\Delta_{\rm H}$.} The experimental data are measured at dimensionless time $\tau$ (in units of $\hbar/t_x$) for $t_y=3.39t_x$ and $U=6.56t_x$, after applying an instantaneous tilt $E_x=0.5t_x$. The values of $J_x$ and $P_y$ (upper plot, respectively red and green) are evaluated by averaging two individual sets of measurements for $\varphi=+0.32\pi$ and $\varphi=-0.32\pi$, each comprising $10\sim15$ images at every time step; the error bars represent standard error of mean and are obtained with a statistical Bootstrap method. The values of $\Delta_{\rm H}$ (lower plot, blue) are computed from the data in the upper plot according to Eq.~\eqref{eq1}, and the error bars represent standard error of mean and are obtained with standard uncertainty propagation. The shaded areas are theoretical predictions accounting for the distribution of atom numbers in the tubes and experimental temperature uncertainty $1\leq T/t_x\leq 2$. They result from a mean-field approximation (MFA, see main text), where the renormalized tunneling $t_y^*=5t_x$ is evaluated through comparison with zero temperature DMRG. The parameter $\widetilde{t_y}$, is introduced to allow meaningful comparison between MFA and experiment. The gray dashed line indicates the universal relation~Eq.~\eqref{eq5}.}
	\label{fig2}
\end{figure}

To generate a current along $\hat{x}$, we switched on an optical gradient $H_{\rm quench}=-E_x\sum_{j,m}j n_{j,m}$, tilting the ladder along the real-lattice direction, with $E_x=0.5t_x$. After time $\tau$, we measured the particle current $J_x$ in the real dimension and the spin polarization $P_y$ in the synthetic dimension. To perform these measurements, the Raman coupling was abruptly switched off to freeze the population along the synthetic dimension. The lattice momentum distribution in the real dimension  $n(k,\tau)$, normalized to the total atom number, was then measured with a band-mapping technique, where the lattice momenta $k$ are expressed in units of the real-lattice wave number $k_{\rm L}=\pi/d$ and $d=\SI{380}{nm}$ is the lattice spacing. We thus access the current $J_x$, given by
\begin{equation}\label{eq3}
	J_x(\tau)=\int^{1}_{-1}\sin(\pi k)n(k,\tau)\,dk
\end{equation}
In the synthetic dimension, the spin distribution is measured by performing an optical Stern-Gerlach detection~\cite{PhysRevLett.105.190401}. This method, based on the spin-dependent force exerted by a near-resonant laser beam, allows a spatial separation of the two spin components and the separate count of the atom number $N_m$ in both of them. The spin polarization coincides with the transverse (Hall) polarization $P_y$ of the system, which we define as
\begin{equation}\label{eq4}
	P_y(\tau)=\frac{N_{1}(\tau)-N_{2}(\tau)}{N_{1}(\tau)+N_{2}(\tau)}-\frac{N_{1}(0)-N_{2}(0)}{N_{1}(0)+N_{2}(0)}
\end{equation}
This definition evaluates the difference in fractional spin population with respect to the initial value, with the populations $N_{1}(0)$ and $N_{2}(0)$ measured right before the application of the optical gradient. The definition in Eq.~\eqref{eq4} accounts for the small initial population difference caused by residual off-resonant coupling to the nuclear spin state $\lvert m_{\rm F}=+3/2\rangle$~\cite{SS}; this initial difference can safely be neglected owing to the averaging procedure discussed in the next section. We determined the Hall imbalance $\Delta_{\rm H}$ from the ratio between the measured $P_y$ and $J_x$, following Eq.~\eqref{eq1}.

\subsection*{Measuring the Hall effect}

Figure~\ref{fig2} shows the measured current, polarization, and Hall imbalance as a function of time $\tau$ (defined in units of $\hbar/t_x$, where $\hbar$ is the reduced Planck's constant) for a particular choice of experimental parameters $t_y=3.39t_x$ and $U=6.56t_x$. We performed identical measurements with both $\varphi=+0.32\pi$ and a reversed direction of the synthetic magnetic field $\varphi=-0.32\pi$ and observed a change of the sign in $P_y(\tau)$~\cite{SS}. This behavior confirms the interpretation of our data in terms of the emergence of a Hall response. We averaged these two independent measurements of \{$J_x(\tau)$, $P_y(\tau)$\} for $\varphi=+0.32\pi$ and \{$J_x(\tau)$, $-P_y(\tau)$\} for $\varphi=-0.32\pi$ to improve the signal-to-noise ratio and minimize the effect of the residual off-resonant coupling to the third state~\cite{SS}. We observed that the Hall imbalance $\Delta_{\rm H}$ (Fig.~\ref{fig2}, bottom) rapidly approaches a stationary regime, with small amplitude deviations around a limiting value, while the dynamical evolution of $J_x$ and $P_y$ remains transient. This fast convergence of $\Delta_{\rm H}$ is reproduced by the theoretical model described in the next paragraph and conveniently allows us to measure the stationary Hall response using quench dynamics.

According to the theoretical predictions reported in~\cite{PhysRevLett.122.083402}, the stationary Hall imbalance for strong interactions ($U\gg t_x$) is expected to reach the $U$-independent universal value
\begin{equation}\label{eq5}
	\Delta_{\rm H}=2\frac{t_x}{t_y}\left|\tan\left(\frac{\varphi}{2}\right)\right|
\end{equation}
Our simulator yields results consistent with this universal value (Fig.~\ref{fig2}, bottom) despite important differences with the setup used in~\cite{PhysRevLett.122.083402} (parabolic confinement, non-linear drive, tubes with different particle occupations and finite temperatures $T\simeq t_x$) and without using any fitting procedure. To explain this robustness, we provide a theoretical analysis. First, we performed extensive density matrix renormalization group (DMRG)~\cite{PhysRevLett.69.2863} simulations at zero temperature (finite-temperature DMRG would be prohibitively costly). To give a semi-quantitative account of the effects of finite temperatures in the non-universal regime (intermediary $U$ and $t_y$), we resorted to a mean-field approximation (MFA) of interactions, which predicts that interactions lead to an effective increase of the transverse coupling $t_y\rightarrow t_y^*$. For each value of $U$, we first find the $t_y^*$ that best reproduces the zero-temperature DMRG real-time simulations of the current $J_x$ and polarization $P_y$. We find a quantitative agreement between MFA and DMRG, if the MFA polarization is multiplied by $t^*_y/t_y$; no rescaling is required for $J_x$. We discuss the clear limitations of MFA to describe the dynamics of such strongly correlated low-dimensional systems~\cite{SS}. Nonetheless, this approach gives a reasonable account of the reason why, at finite temperatures, larger values of the transverse hopping $t_y$ or the interaction $U$ are required to observe the universal Hall response~\eqref{eq5}, as observed and explained below. 

\begin{figure}[t!]
	\centering
	\includegraphics[width=\columnwidth]{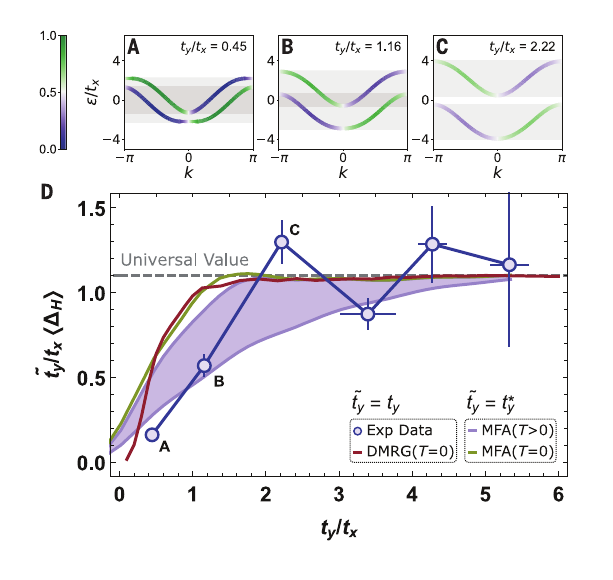}
	\caption{\textbf{Time-averaged Hall imbalance as a function of synthetic tunneling.} \textbf{(A-C)} Single-particle energy spectrum $\varepsilon(k)$ calculated as a function of the quasimomentum $k$ for different values of $t_y/t_x$ (A, B, C in panel (D)): The interband gap increases with $t_y/t_x$, eventually leading to two separate bands (C). The color scale represents the population of the $m=1$ state. \textbf{(D)} The experimental data (blue circles) are measured at $U=6.56t_x$ and $|\varphi|=0.32\pi$, with the averaging procedure and error analysis detailed in Fig.~\ref{fig2}. The horizontal and vertical error bars show the experimental uncertainty in $t_y$ and the uncertainty resulting from the time average, respectively. The red solid line is the DMRG simulation at zero temperature for a fixed atomic interaction $U=6.56t_x$ and a number of rungs $L=200$, accounting for different tube occupations. The yellow solid line is the MFA at zero temperature with renormalized $t^*_y=t_y+0.1U$ (see main text for details). The shaded area illustrates the MFA  at finite temperatures $0.5\leq T/t_x\leq 2$, and the gray dashed line indicates the universal relation from Eq.~\eqref{eq5}.}
	\label{fig3}
\end{figure}

\subsection*{Testing universality}

To pinpoint the onset of the universal regime, we measured the dependence of the Hall imbalance's stationary value on the system parameters. We considered the averaged Hall imbalance $\langle\Delta_{\rm H}\rangle=\langle P_y(\tau)/J_x(\tau)\rangle_\tau$ in the time interval $\tau\in[1,5]$. Figure~\ref{fig3}D shows the measured $\langle\Delta_{\rm H}\rangle$ for a fixed interaction strength $U=6.56t_x$ and different values of the tunneling ratio $t_y/t_x$, which is controlled by changing the Raman beam power. The averaged Hall imbalance $\langle\Delta_{\rm H}\rangle$ is small at small synthetic tunneling ($t_y\ll t_x$) and reaches the universal value~\eqref{eq5} for $t_y/t_x\gtrsim 2$. Provided that the system is below half-filling, this transition also exists in non-interacting systems, where a large transverse hopping $t_y$ opens a large gap between the two bands of the system, stabilizing a single-band metal whose Hall imbalance has the universal value~\eqref{eq5}~\cite{PhysRevLett.122.083402,SS} (see single-particle energy bands in Fig.~\ref{fig3}, A to C).
Since finite temperatures tend to promote particles to the upper band, we expect the finite temperature ($T\simeq t_x$) in our setup to push the transition to the universal regime towards larger values of $t_y/t_x$. This effect is visible in Fig.~\ref{fig3}: Whereas zero-temperature DMRG predicts a transition to the universal regime at smaller values of $t_y/t_x$, the finite-temperature MFA yields a better quantitative agreement with the experiment.

\begin{figure}[t!]
	\centering
	\includegraphics[width=\columnwidth]{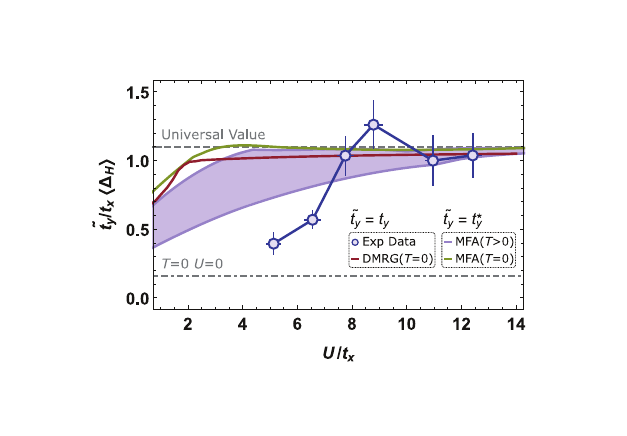}
	\caption{\textbf{Time-averaged Hall imbalance as a function of atomic interaction.} The experimental data (blue circles) are measured at $t_y=1.15t_x$ and $|\varphi|=0.32\pi$, with the averaging procedure and error analysis detailed in Fig.~\ref{fig2}. The horizontal and vertical error bars show the experimental uncertainty in $U$ and the uncertainty resulting from the time average, respectively. The red solid line is the DMRG simulation at zero temperature for a fixed synthetic tunneling $t_y=1.15t_x$ and $L=200$ rungs, accounting for different tube occupations. The yellow solid line is the MFA at zero temperature, with the substitution $t_y\rightarrow t_y^*=t_y+0.30\cdot U$. The shaded area illustrates MFA results for finite temperatures $0.5\leq T/t_x\leq 2$. The gray dashed line indicates the universal relation from Eq.~\eqref{eq5}, and the dot-dashed line depicts the result for non-interacting fermions at zero temperature.}
	\label{fig4}
\end{figure}

Finally, we demonstrate the interaction-driven origin of the universal Hall response of Eq.~\eqref{eq5}. Figure~\ref{fig4} illustrates the behavior of the Hall imbalance $\langle\Delta_{\rm H}\rangle$ upon changing the interaction strength $U/t_x$ at a fixed, nearly isotropic tunneling $t_y=1.15t_x$. 
We observe that $\langle \Delta_{\rm H}\rangle$  quickly deviates from the non-interacting value and approaches the $U$-independent universal value $2\left(t_x/t_y\right)\left|\tan\left(\varphi/2\right)\right|\simeq 1.1$ at large $U/t_x$. 
In the spirit of the MFA, this behavior can be partially explained by the two-band scenario discussed before: Interactions renormalize $t_y$ towards an interaction-dependent value $t_y^*>t_y$, enlarging the gap between the bands. In the large-$U$ limit, the bands are well separated and the highest band becomes empty, similarly to what happens in the large-$t_y$ limit (Fig.~\ref{fig3}, A to C). Increasing $U$ thus leads to a robust single-band metallic state characterized by a constant, universal value of $\Delta_{\rm H}$~\cite{PhysRevLett.122.083402}.
As discussed above, the MFA accounts for finite-temperature effects and permits a quantitative comparison between experiment and theory. Similar to the transition from weak to large transverse tunneling described in the previous paragraph, the finite temperature pushes the transition towards larger interaction strengths than those predicted by the zero-temperature DMRG, as observed in the experimental data. 

Despite the effectiveness of the MFA picture, we stress the essential and non-perturbative role played by strong interactions to reach the universal regime in Fig.~\ref{fig4}, which fundamentally differentiates it from the large-$t_y$ limit (Fig.~\ref{fig3}). Indeed, while strong interactions can be modelled by the MFA using a renormalized $t_y^*$, the universal regime is reached anyway for $\Delta_{\rm H}=2\left(t_x/t_y\right)\left|\tan\left(\varphi/2\right)\right|$, and not for $\Delta_{\rm H}=2\left(t_x/t_y^*\right)\left|\tan\left(\varphi/2\right)\right|$. We emphasize that the observed effect is truly a many-body effect, as captured by the DMRG calculations. The MFA allows us to make progress at finite temperatures because it reproduces the stabilization of the single-band metal at large $U$ by increasing $t_y$, where the Hall imbalance approaches a universal value when $t_y>T$~\cite{SS}. Nonetheless, MFA requires its results to be rescaled to give a quantitative account of the universal regime and has clear limitations with respect to reproducing the fine dynamics of the polarization at large $U$~\cite{SS}. The MFA should thus be complemented with more complete, but much more difficult, exact finite-temperature studies.

\subsection*{Conclusions}

In this experiment, we have shown distinctive many-body effects triggered by strong interactions in the Hall response of a controllable quantum simulator of a two-leg ladder threaded by a synthetic magnetic flux. Beyond the clear potential of such experiments to measure Hall voltages~\cite{buser2021} and clarify the exotic Hall response of strongly correlated solid-state conductors, our work paves the way to the investigation of the exotic transport properties of strongly correlated topological phases of matter. This cold-atom experiment enters unknown territory for theory because it features strong correlations and finite temperatures and yet shows full control of the simulation parameters. An interesting perspective resides in investigating interacting ladders with a larger number of nuclear spin states, a regime notoriously difficult to access with present computational techniques.

\paragraph*{Acknowledgments}
We thank T. Beller for very helpful comments on the manuscript and C. Berthod and N. Cooper for useful discussions.

\paragraph*{Funding}
We acknowledge financial support from the Topology and Symmetries in Synthetic Fermionic Systems (TOPSIM) European Research Council (ERC) Consolidator Grant 682629, the TOPSPACE MIUR FARE project, Quantum Technologies For LAttice Gauge (QTFLAG) QuantERA ERA-NET Cofund in Quantum Technologies, and MIUR PRIN project 2017E44HRF. This work was supported in part by the Swiss National Science Foundation under Division II grant 200020-188687. M.F. acknowledges support from Swiss National Science Foundation/Schweizerischer Nationalfonds (FNS/SNF) Ambizione grant PZ00P2\_174038 and the EPiQ ANR-22-PETQ-0007 part of Plan France 2030.

\paragraph*{Author contributions}
L.Fa., J.C., G.C., M.I., M.F., and T.G. conceived the experiments. T.-W.Z., D.T., L.Fr., G.C., and J.P. carried out the experimental work. T.-W.Z., D.T., and L.Fr. analyzed the experimental results. S.G., C.R., M.F., and T.G. performed theoretical work. All authors contributed extensively to the discussion of the results and to the writing of the manuscript.

\paragraph*{Competing interests}
The authors declare no competing interests.

\paragraph*{Data and material availability}
DMRG and time-dependent variational principle calculations were performed using the TeNPy Library (version 0.7.2)~\cite{tenpy1,tenpy2}. All of the experimental and theoretical data presented in the main figures are available for download from zenodo, an open-access repository~\cite{Zenodo}.




%


\newpage
\clearpage

\renewcommand{\thefigure}{S\arabic{figure}}
 \setcounter{figure}{0}
\renewcommand{\theequation}{S.\arabic{equation}}
 \setcounter{equation}{0}
\renewcommand{\thesection}{S.\Roman{section}}
\setcounter{section}{0}
\renewcommand{\thetable}{S\arabic{table}}
 \setcounter{table}{0}
\renewcommand{\bibnumfmt}[1]{[S#1]}
\renewcommand{\citenumfont}[1]{S#1}

\onecolumngrid

\begin{center}
{\bf \large Supplementary Material for\\
\vspace{3mm}
``Observation of Universal Hall Response in Strongly Interacting Fermions''}\\
\vspace{3mm}
T.-W. Zhou, G. Cappellini, D. Tusi, L. Franchi, J. Parravicini, C. Repellin,\\
S. Greschner, M. Inguscio, T. Giamarchi, M. Filippone, J. Catani, L. Fallani
\end{center}

\twocolumngrid

\section{Adiabatic state preparation}

\begin{figure}[t!]
    \centering
    \includegraphics[width=0.6\columnwidth]{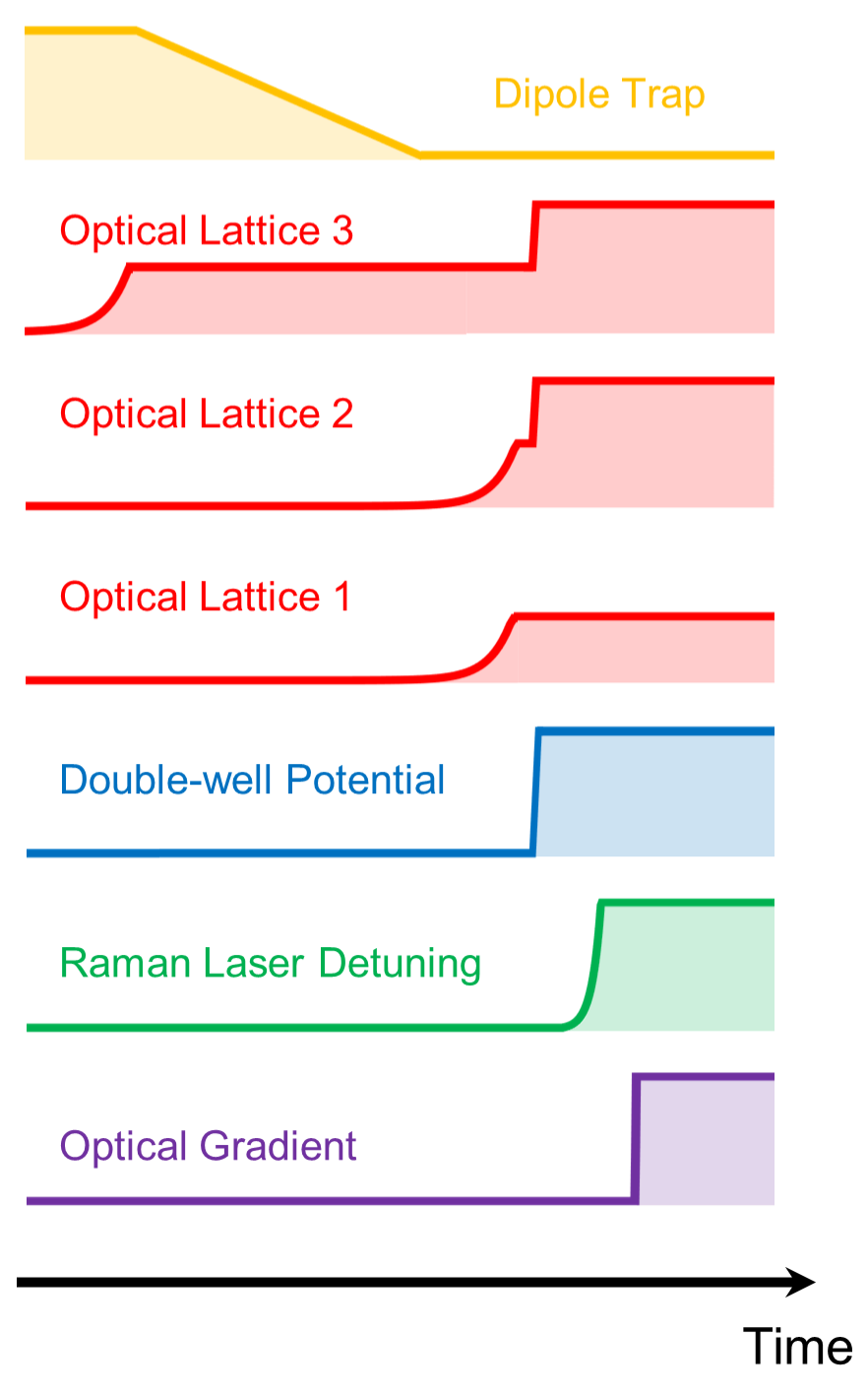}
    \caption{Sketch of experimental procedure sequence. See text for details.}
    \label{figS1}
\end{figure}

The $^{173}$Yb spin-polarized degenerate Fermi gas ($\lvert m_{\rm F}=-5/2\rangle$) with a typical temperature of $0.2T_{\rm F}$, where $T_{\rm F}$ is the Fermi temperature, is loaded into the vertical lattice (Optical Lattice 3, along the direction of gravity) within $\SI{150}{ms}$ using an exponential intensity ramp. The optical dipole trap, in which the atoms are originally confined, is subsequently switched off in $\SI{1}{s}$. Two horizontal lattices (Optical Lattice 1\&2, see Fig.~\ref{figS1}) are then ramped up in the same way as the vertical lattice. The three lattice depths are set to $V_{\rm{OL1}}=4$ or $5\ E_{\rm r}$, and $V_{\rm{OL2}}=V_{\rm{OL3}}=15\ E_{\rm r}$, where $E_{\rm r}=h^2/8m d^2$ is the recoil energy, $h$ is the Planck constant and $m$ is the atomic mass. The tunneling rate $t_x/\hbar$ (either $2\pi\times \SI{171}{Hz}$ or $2\pi\times \SI{131}{Hz}$) along the fermionic tubes is thus much larger than the radial tunneling rates ($\sim\SI{12.96}{Hz}$), which ensures the dynamics is only allowed in the shallow lattice along longitudinal direction $\hat{x}$. These independent 1D fermionic tubes are characterized by an axial harmonic confinement with a trapping frequency of $2\pi\times\SI{43}{Hz}$, which originates from the Gaussian intensity profiles of the 2D red-detuned lattice beams.

In order to keep the same atom number distribution among the fermionic tubes for different radial confinement strength, the 2D lattice is always adiabatically ramped up to $15\ E_{\rm r}$, followed by a rapid linear ramp to a final lattice depth in $\SI{1}{ms}$, which is completely negligible compared with the inter-tube tunneling time ($\sim\SI{77.16}{ms}$). The optical dipole potential compensating the longitudinal trapping frequency (see Section S.IV. for details) is also switched on linearly in $\SI{1}{ms}$ during the same time window of the rapid linear ramp of the 2D lattice, so as to avoid excitations in the longitudinal direction.

After the lattice loading procedure, we switch on the Raman laser beams with an initial detuning $\delta_{\rm i}=-5\sim\SI{-15}{kHz}$ and perform an exponential frequency sweep of the form
\begin{equation}
	\label{delta_t}
	\delta(t)=\delta_{\rm i}+(\delta_{\rm f}-\delta_{\rm i}) \left(\frac{1-e^{-t/T_{\rm{tau}}}}{1-e^{-T_{\rm{adiab}}/T_{\rm{tau}}}}\right)\,,
\end{equation}
where $\delta_{\rm f}$ is chosen to resonantly couple the two nuclear spin states $\lvert m_{\rm F}=-1/2\rangle$ and $\lvert m_{\rm F}=-5/2\rangle$. The ramp duration $T_{\rm{adiab}}$ ranges from $20\sim\SI{60}{ms}$ depending on the experimental configuration, with $T_{\rm{tau}}$ ranging from $5\sim\SI{20}{ms}$ accordingly. The adiabaticity of the whole process is verified experimentally by reversing the whole procedure to recover a spin-polarized Fermi gas.

\section{Spin-resolved Detection}

After we suddenly switch off the Raman coupling, thus freezing the population along the synthetic direction $\hat{y}$, all the lattice potentials are exponentially ramped down in $\SI{1.2}{ms}$. To measure the Hall polarization $P_y$, the two spin components are separately imaged by exploiting an optical Stern-Gerlach technique~\cite{sPhysRevLett.105.190401}. This spin-resolved measurement is implemented by shining a $\SI{10}{mW}$ circularly polarized laser pulse with a red-detuning of $\SI{866}{MHz}$ from the $^1S_0\rightarrow^3P_1\ (F'=7/2)$ transition during the first $\SI{1.5}{ms}$ of the time-of-flight (TOF) period.

\section{Optical gradient}

\begin{figure}[t!]
    \centering
    \includegraphics[width=\columnwidth]{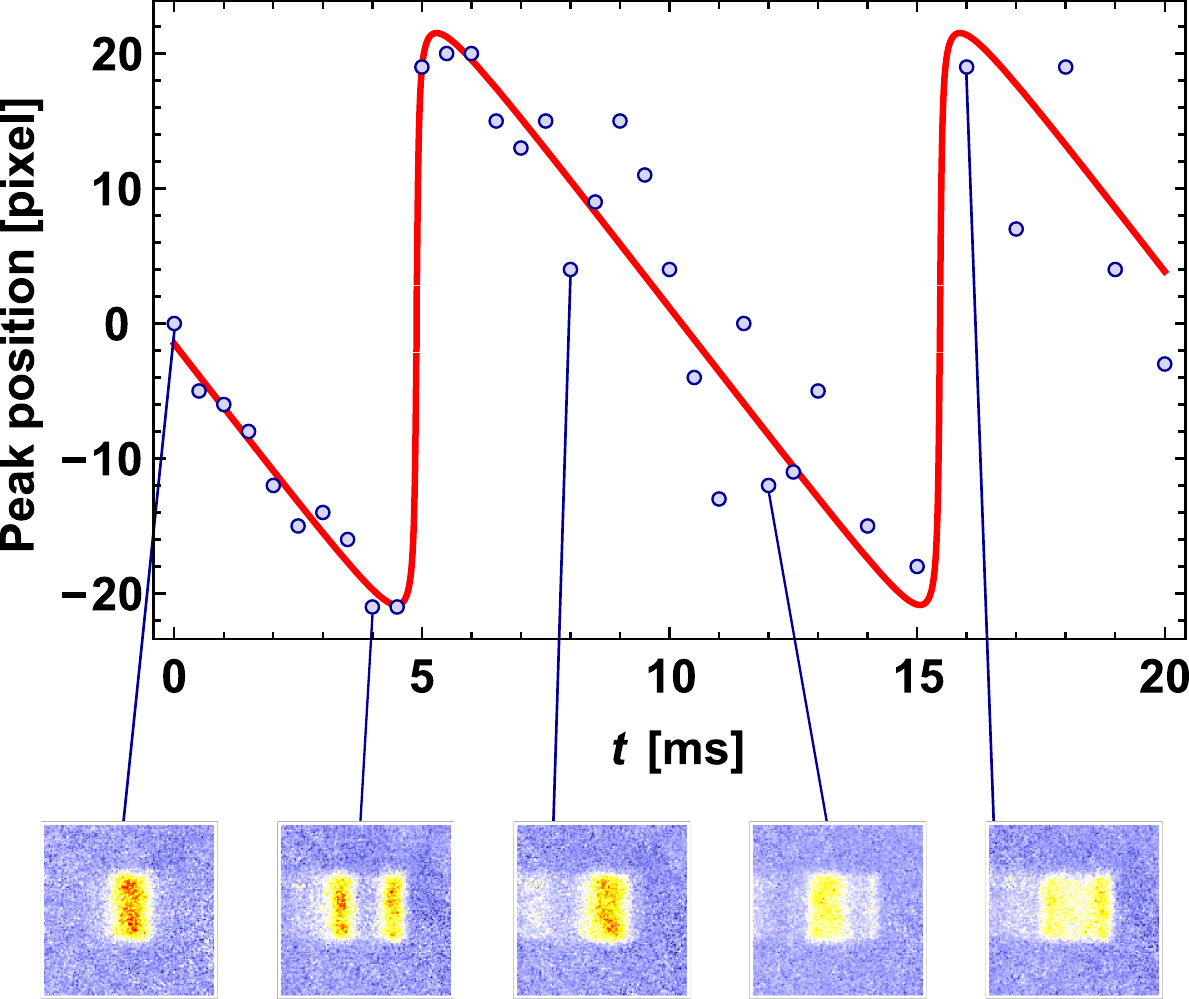}
    \caption{Typical experimental Bloch oscillation. Position of the peak in the momentum distribution as a function of lattice hold time $t$. The red solid line is a fit to the data yielding a gradient strength around $E_x=0.5t_x$. Experimental TOF images of the Bloch oscillation at $t=0,\,4,\,8,\,12,\,\SI{16}{ms}$ are shown in the lower panel.}
    \label{figS2}
\end{figure}

The optical gradient used to induce the current along the $\hat{x}$ direction is realized by a focused laser beam operating at $\SI{1112}{nm}$ with the atomic cloud center located at the maximum slope of the Gaussian beam. The exact value of the optical gradient is determined by a separate Bloch oscillation measurement as shown in Fig.~\ref{figS2}, with a spin-polarized Fermi gas in optical lattices at potential depth $V_{\rm{OL1}}=1.2\ E_{\rm r}$ and $V_{\rm{OL2}}=V_{\rm{OL3}}=12\ E_{\rm r}$.

\section{Trapping frequency compensation}

As mentioned in the main text, the ``on-rung'' interaction energy $U$ is controlled by tuning the 2D lattice depth. A subsequent issue is that the longitudinal confinement strength of the fermionic tubes will change as well. We overcome this by superposing a double-well potential with adjustable barriers along direction $\hat{x}$, through which the overall longitudinal trapping frequency can be well adapted for different experimental configurations.

\begin{figure}[t!]
    \centering
    \includegraphics[width=\columnwidth]{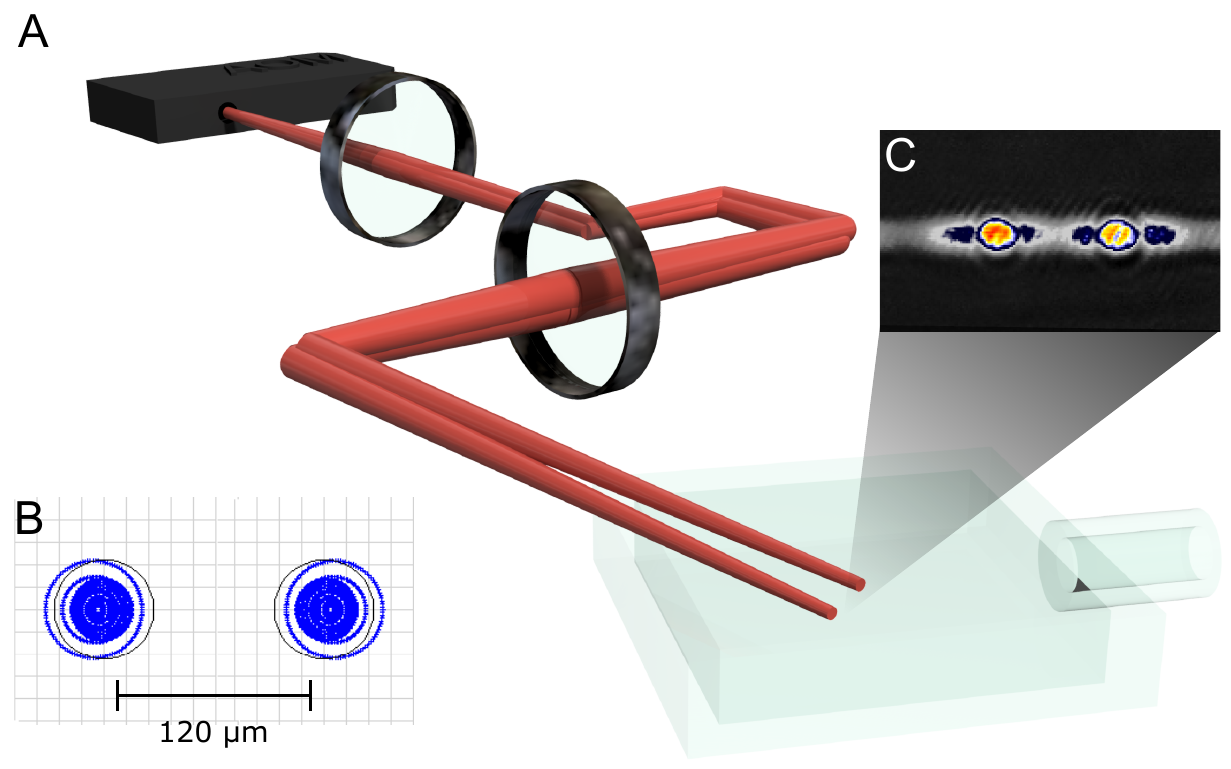}
    \caption{Simplified sketch of the double-well setup designed to compensate the lattice residual harmonic confinement. (\textbf{A}) Schematics of the setup compensating the longitudinal trapping frequency. A double-well potential is imaged onto the atom clouds realizing an anti-trapping potential along longitudinal direction $\hat{x}$. (\textbf{B}) Zemax OpticStudio$^{\copyright}$ simulation. Black circles represent the Airy disks due to diffraction limit. (\textbf{C}) False-color experimental TOF images of the atoms in the double-well potential (averages of $\sim10$ realizations).}
    \label{figS3}
\end{figure}

To achieve this, similar to what was done in Ref.~\cite{sPhysRevLett.92.050405}, we shine a collimated laser beam ($\SI{1112}{nm}$) onto an acousto-optic modulator (AOM) which is driven by two radio-frequency (RF) signals ($\SI{107.6}{MHz}$ and $\SI{112.4}{MHz}$, respectively). The two diffracted beams are then focused with a lens (focal length $\SI{100}{mm}$) to obtain a focused beam waist around $\SI{70}{\mu m}$. The AOM is placed in the focal plane of the lens to keep the two beams propagating parallel to each other, and the distance between the two potential wells ($\sim\SI{120}{\mu m}$) is determined by the frequency difference between the two RF signals. This double-well potential is final imaged onto the atom clouds from a direction perpendicular to direction $\hat{x}$, as depicted in Fig.~\ref{figS3}A. The main point is to include the potential barrier between the two wells, which has an anti-trapping effect to compensate the trapping frequency, along the 1D fermionic tubes. While the measurements of interaction effect on Hall response are carried out at different 2D lattice depths, the light intensity of the double-well potential is changed accordingly to adjust the barrier in the middle, and keep the longitudinal confining strength $V_x=0.01t_x$ unchanged for all different $U/t_x$.

\section{Weak off-resonant coupling to the third state}

\begin{figure}[t!]
	\centering
	\includegraphics[width=\columnwidth]{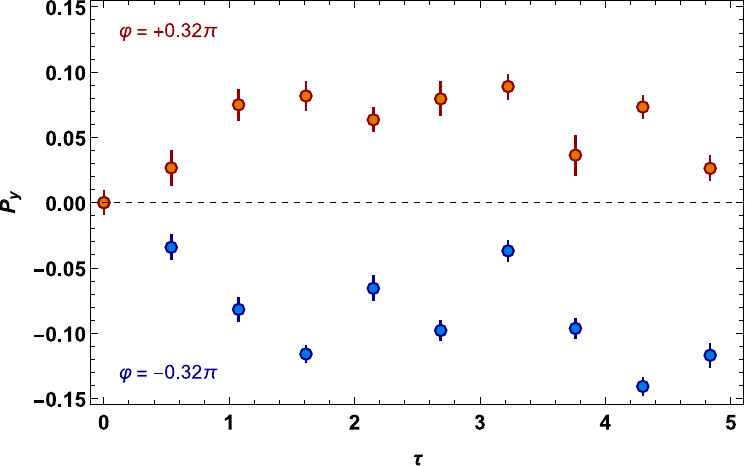}
	\caption{The time evolution of Hall polarization $P_y$ for opposite directions of synthetic magnetic field $\varphi=\pm 0.32\pi$. The data points are measured at $t_y=1.15t_x$ and $U=7.76t_x$. The error bars represent standard error of mean and are obtained with a statistical Bootstrap method.}
	\label{figS4}
\end{figure}

The different internal states of $^{173}$Yb are coupled by a two-photon Raman transition, providing coherent controllable couplings between different spin components. When the Raman laser detuning makes the $\lvert m_{\rm F}=-5/2\rangle\to\lvert m_{\rm F}=-1/2\rangle$ transition perfectly resonant, the Hamiltonian that describes the Raman coupling in our experimental system can be expressed as 
\begin{equation}
	\hat{H}_{\rm R} = t_y
	\begin{pmatrix}
		0 & 1 & 0 \\
		1 & 0 & \alpha \\
		0 & \alpha & 2\beta \\
	\end{pmatrix}\,,
\end{equation}
in the rotating frame after adiabatic elimination of the excited states, where $\alpha=1.41$ and $\beta=2.65$. Under this condition, the population of the third state $\lvert m_{\rm F}=+3/2\rangle$ which is only weakly coupled to the other two spin states, is at most a few percent and thus negligible throughout the experiment. However, this weak off-resonant coupling does have an effect on the following Hall dynamics when the instantaneous quench of the linear potential occurs, particularly on the quantity of $P_y$. This is verified in our experiment (see Fig.~\ref{figS4}) and further confirmed by the results of the non-interacting simulations.

\begin{figure*}[t!]
    \centering
    \includegraphics[width=2\columnwidth]{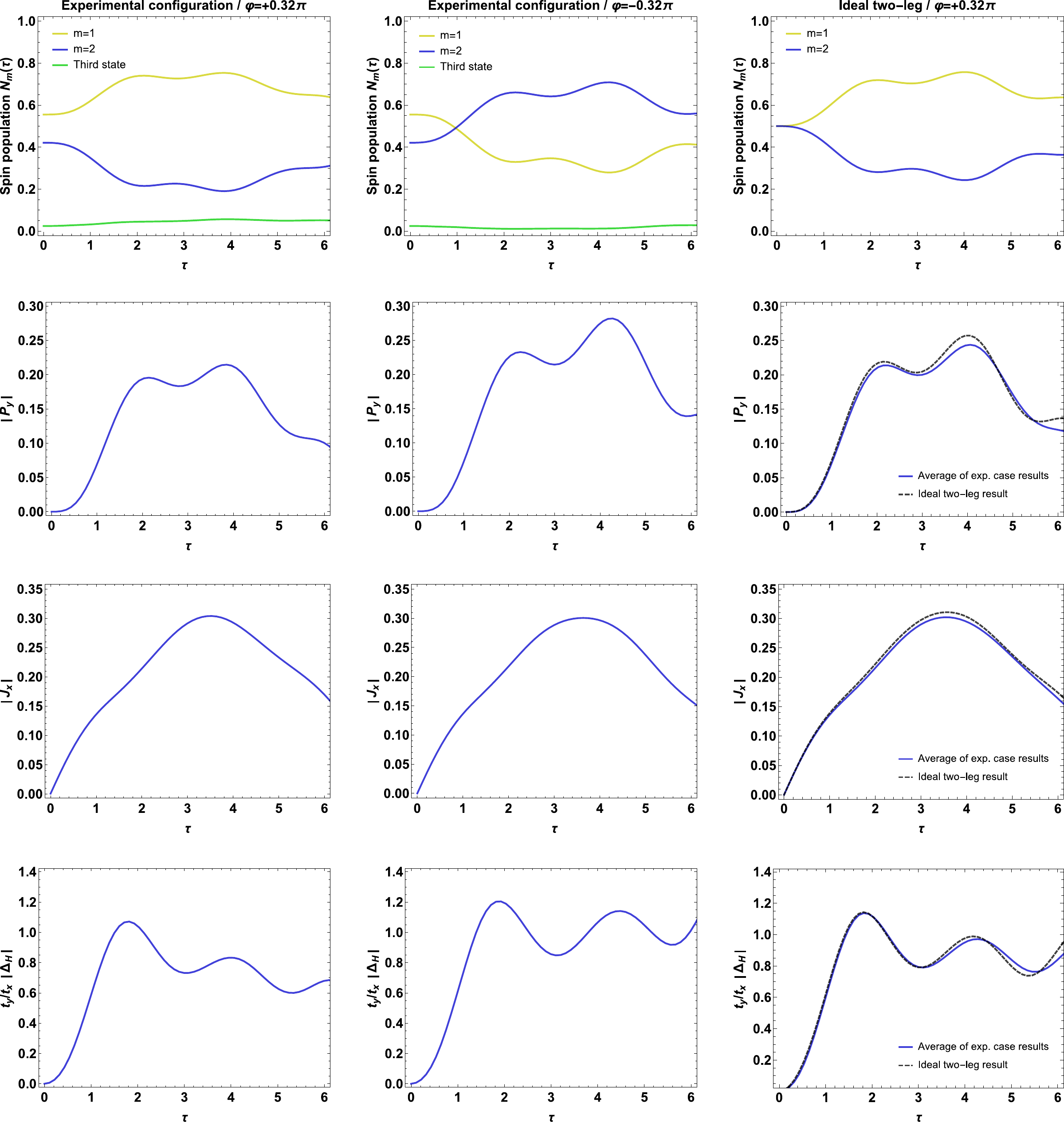}
    \caption{Non-interacting simulation results ($N=10$, $L=100$) for $t_y=1.15t_x$ and $V_x=0.01t_x$, following a linear potential quench of strength $\mu_x=0.5t_x$. First two columns: results of the experimental configuration (weak off-resonant coupling to the third state) for $\varphi=+0.32\pi$ and $\varphi=-0.32\pi$. Third column: results of an ideal two-leg case for $\varphi=+0.32\pi$, together with the average of the results of the experimental case for $\varphi=+0.32\pi$ and $\varphi=-0.32\pi$.}
    \label{figS5}
\end{figure*}

Figure~\ref{figS5} shows a non-interacting simulation to illustrate the effect of the third state weak coupling and how we get rid of it by an averaging procedure. The first two columns show the results of the experimental configuration taking into account weak off-resonant coupling to the third state, for $\varphi=+0.32\pi$ and $\varphi=-0.32\pi$; the third column shows the results of an ideal two-leg, together with the average of the results of the experimental case for $\varphi=+0.32\pi$ and $\varphi=-0.32\pi$. As we can see from Fig.~\ref{figS5}, averaging the results for $\varphi=+0.32\pi$ and $\varphi=-0.32\pi$, with the definition~(4) of $P_y$, brings a substantial agreement with what is  expected in the pure two-leg case (blue solid vs black dashed lines in the last plots of rows 2-4), thus neutralizes the effect of weak coupling to the third state. There are some deviations at large times, though, when the population of the third state increases, which is not problematic since we only focus on the transient dynamics before the third state population starts to have sensitive effects on the system dynamics.

\section{Number of atoms in fermionic tubes}\label{sec:no_of_atoms}

As it is mentioned in the main text, the resulting potential of two deep orthogonal lattices consists of an array of independent 1D tubes with residual harmonic confinement. Arising from the Gaussian intensity profiles of the red-detuned lattice beams, there is a spatially dependent potential offset for each tube with respect to the central one. The number of atoms contained in each tube is not uniform, instead, the larger this potential offset, the less the number of particles required to fill the tube up to the Fermi energy of the system.

\begin{figure}[t!]
    \centering
    \includegraphics[width=\columnwidth]{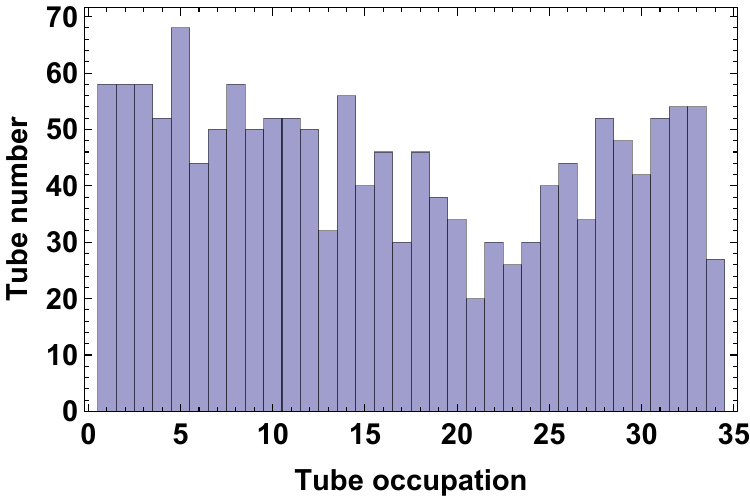}
    \caption{Number of atoms in fermionic tubes for the typical total atom number $N_{\rm atom}\simeq 2.5\times10^4$.}
    \label{figS6}
\end{figure}

The number of atoms in different tubes is estimated from the measured trapping frequencies and the total atom number $N_{\rm atom}$. We perform a numerical zero-temperature simulation where we treat the lattice confining potentials as the sum of two harmonic potentials (radial direction, single-site contributions coming from harmonic potential energy) and a shallow lattice with residual harmonic confinement (axial direction, eigenstate energies obtained from exact diagonalization calculation). Starting from the center region with the lowest energy, each lattice site is filled with at most one atom due to the Fermi statistic, until $N_{\rm atom}$ is reached.

It is worth noting that the atom number in tubes has already been determined before we initiate the Raman coupling, thus we only need to consider free fermions in this estimation. From result of the simulation we can easily obtain the atom number distribution among the 2D array of tubes. For the typical total atom number $N_{\rm atom}\simeq 2.5\times10^4$ in the experiment, atoms are distributed in $\sim 1525$ tubes with a central tube occupation of $34$ (see Fig.~\ref{figS6}).

\section{Estimation of Temperature in lattices}

In this section, we give a temperature estimation of the system through a thermodynamics approach, where the local-density approximation (LDA) is applied to a grand-canonical ensemble. The properties of Raman coupled two-component fermions trapped in optical lattices are derived, based on the grand-canonical partition function $Z(T,\mu)$ (where $T$ is the temperature in the lattice and $\mu$ is the global chemical potential) and free energy $F(T,\mu)=-k_{\rm B}T\ln{Z(T,\mu)}$, where $k_{\rm B}$ is the Boltzmann constant. Here, we assume that the lattice loading procedure is isentropic, meaning the entropy $S$ is fixed to the value of entropy in the harmonic dipole trap before lattice loading. For a non-interacting Fermi gas in harmonic trap, the expression for the entropy is $S=k_{\rm B}N_{\rm atom}\pi^2T_{\rm{Trap}}/T_{\rm F}$~\cite{sPhysRevLett.92.150404}, where $N_{\rm atom}$ is the total atom number, and the ratio between the temperature in the trap
and the Fermi temperature $T_{\rm{Trap}}/T_{\rm F}=0.2$ are both experimentally accessible quantities.

We then apply LDA in which the system is considered locally at equilibrium~\cite{sHo2009}. Thus, the grand-canonical free energy can be written as $F(T,\mu)=\sum_{j}F(T,\mu-V_j)$, where $V_j$ is the external potential for a given lattice site $j$. Finally we have two coupled equations, from the solution of which we can obtain the temperature in the lattice 
\begin{equation}
	\sum_{j}n_i(T,\mu-V_j)=\sum_{j}-\frac{\partial F(T,\mu-V_j)}{\partial \mu}=N\\\,,
\end{equation}
\begin{equation}
	\sum_{j}s_i(T,\mu-V_j)=\sum_{j}-\frac{\partial F(T,\mu-V_j)}{\partial T}=S\,.
\end{equation}

For a relatively high temperature, the tunneling between neighbouring lattice sites can be further taken into account by employing a high-temperature series expansion~\cite{sTaie2012}. The calculation results show that the temperature in the lattice is $T\simeq1.0\sim1.5t_x$.

\section{Theoretical analysis}
In this section, we detail the theoretical analysis of our experimental data. We start by giving some insights into the universality of the Hall response~(5) in the $t_y\gg t_x$ limit by analyzing the non-interacting ladder in the linear response regime (\ref{sec:non-interacting}). Using a mean-field approach (\ref{sec:MF}), we then show that repulsive interactions effectively increase the transverse hopping $t_y$, hence reinforcing the robustness of the Hall response. Finally, we give the details of our time-dependent numerical calculations (DMRG and mean-field) with realistic experimental conditions, and show the quantitative agreement between DMRG and mean-field at zero temperature (\ref{sec:realistic time}). This numerical data supports the universal behavior of the Hall response even in the realistic conditions of the experiment.

\subsection{Linear response in the non-interacting ladder}
\label{sec:non-interacting}
\begin{figure}[t!]
	\centering
	\includegraphics[width=\columnwidth]{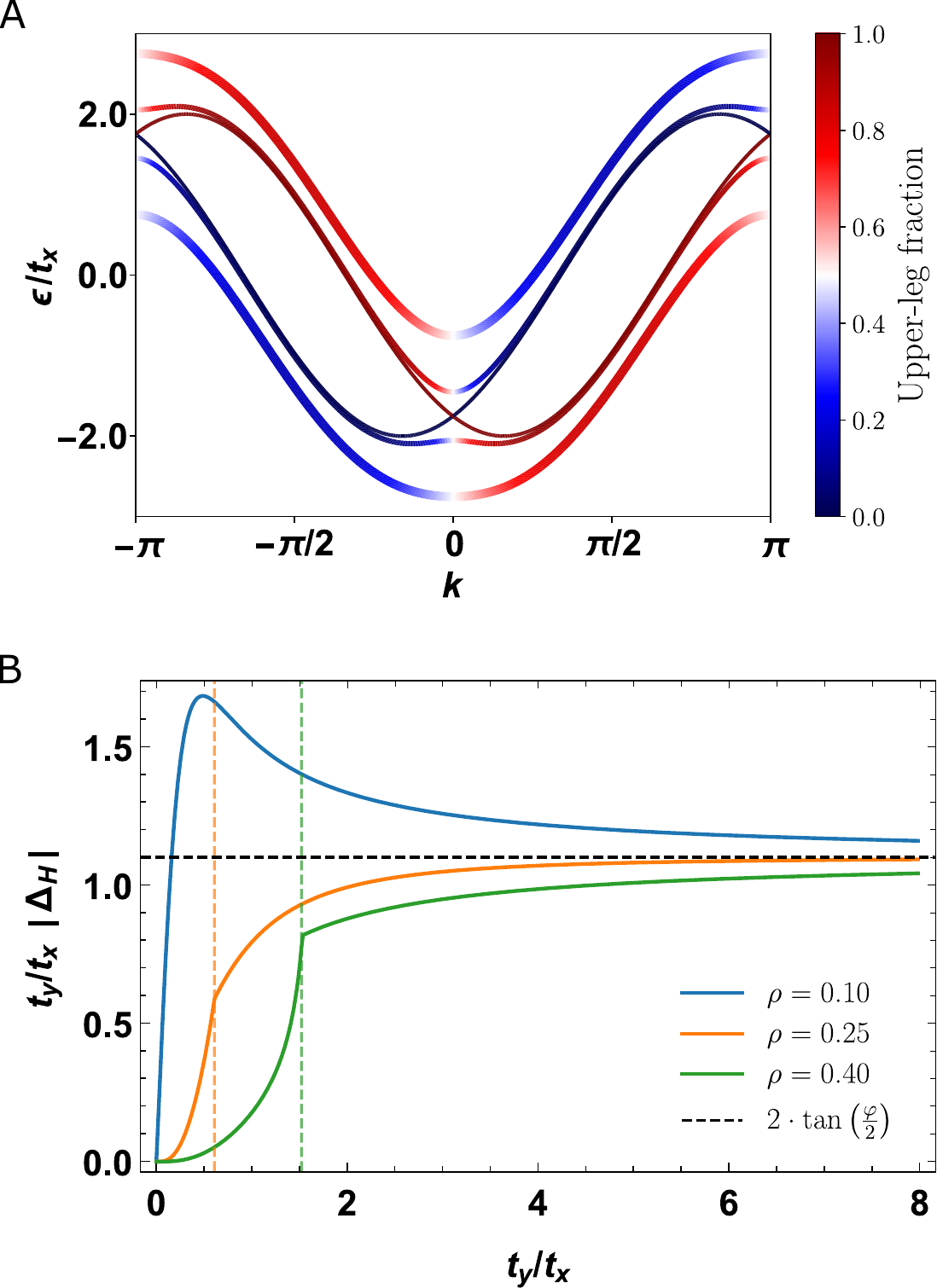}
	\caption{Exact results for the non-interacting ladder at zero temperature. (\textbf{A}) Energy spectrum of the non-interacting ladder (Eq.~\eqref{eq:bands}) as a function of the lattice momentum $k$ and different values of $t_y$ ($t_y/t_x=0,\,0.3$ and 1, thinner to thicker lines). The color on top of the lines indicates the upper-leg fraction of the single particle state. We consider the experimental value of the flux $\varphi=0.32\,\pi$, and different values  of the transverse hopping $t_y$. (\textbf{B}) Hall imbalance $\rm \Delta_{\rm H}$~\eqref{eq:DH} as a function of $t_y/t_x$  for different values of the density $\rho=N/2L$. The vertical dashed lines indicate the transition between two different ground states, whereby the Fermi level crosses both bands or only one.}
	\label{figS7}
\end{figure}

We consider the non-interacting ladder, i.e. the Hamiltonian~(2) with $U=0$ on a ladder with $L$ rungs and periodic boundary conditions $a_{j,m}=a_{j+L,m}$. To calculate the Hall response, we also consider a flux $\phi$ inserted in the ring, which induces a complex hopping term $-t_x\sum_{j,m}(e^{i\phi}a^\dagger_{j,m}a_{j+1,m}+\mbox{h.c.})$, and a chemical potential imbalance $\nu\sum_{j}(n_{j,1}-n_{j,2})$.
We diagonalize the Hamiltonian by switching to reciprocal space; the single-particle spectrum consists of two bands $\varepsilon_{\pm}(k)$
\begin{equation}\label{eq:bands}
	\begin{split}
		\varepsilon_{\pm}(k)=&-2t_x\cos(k+\phi)\cos\left(\frac\varphi2\right)\\
		&\pm\sqrt{\left[2t_x\sin(k+\phi)\sin\left(\frac\varphi2\right)+\nu\right]^2+t_y^2}\,.
	\end{split}
\end{equation}
The spectrum is plotted in Fig.~\ref{figS7}A for the experimental value of the flux $\varphi=0.32\pi$ and different values of $t_y/t_x$ ($\phi = \nu = 0$). At $t_y=0$, the two bands cross at $k=0,\pm\pi$; a finite transverse hopping ($t_y\neq0$) gaps the two bands at the crossings. In Fig.~\ref{figS7}A, we also show the polarization of the single-particle states in the $\pm$-band by showing their upper-leg fraction 
\begin{equation}
	P_\pm(k)=\frac12\left(1\mp\frac{2t_x\sin(k)\sin(\varphi/2)}{\sqrt{t_y^2+4t_x^2\sin^2(k)\sin^2(\varphi/2)}}\right)\,,
\end{equation}
that is their probability to be found in the upper leg. The bands have thus opposite polarizations and, as we are going to illustrate, the Hall response drastically enhances and converges towards the universal regime as soon as the upper band is emptied. 

To derive the Hall response, we follow the approach devised in Refs.~\cite{sPhysRevLett.85.377,sPhysRevLett.83.2785} for the derivation of the Hall constant $R_{\rm H}$, which was adapted in Refs.~\cite{sPhysRevLett.122.083402,sPhysRevLett.123.086803} to address the Hall imbalance $\Delta_{\rm H}$. In the linear response regime, the current $j_x$ and polarization density $p_y$ can be written in terms of the ground state susceptibilities to an infinitesimal flux $\phi$ and chemical potential imbalance $\nu$
\begin{align}\label{eq:jxpy}
	j_x&=\left.\frac1{2t_x L}\frac{\partial\langle\mathcal H\rangle}{\partial \phi}\right|_{\phi,\nu\rightarrow0}\,,& p_y&=\frac1L\left.\frac{\partial \langle\mathcal H\rangle}{\partial \nu}\right|_{\phi,\nu\rightarrow0}\,,
\end{align}
where the $2t_x$ factor in the above definition of the current density is introduced to be consistent with the definition of the Hall imbalance in the main text, Eq.~\eqref{eq1}. 

Equation~\eqref{eq:jxpy} gives the current and the polarization carried by the non-interacting system via derivatives of the band spectrum~\eqref{eq:bands}. As we are interested in the linear response regime, defined by $\phi,\nu\rightarrow0$, we expand the current and the polarization in the longitudinal flux $\phi$ and chemical potential imbalance $\nu$. As a consequence, the Hall imbalance $\Delta_{\rm H}$ in the linear response and non-interacting limit is given by~\cite{sPhysRevLett.122.083402,sPhysRevLett.123.086803}
\begin{equation}\label{eq:DH}
	\Delta_{\rm H}=\left.\frac{p_y}{j_x}\right|_{\phi,\nu\rightarrow0}=2t_x\left.\frac{\sum_{k,s}\partial_{\phi\nu}\varepsilon_{k,s}\cdot n_{k,s}}{\sum_{k,s}\partial_{\phi\phi}\varepsilon_{k,s}\cdot n_{k,s}}\right|_{\phi,\nu\rightarrow0}\,,
\end{equation} 
where $n_{k,s}$ is the probability of occupation of the single particle state labeled by $(k,s)$. For instance, in the grand-canonical ensemble, $n_{k,s}$ is the Fermi-Dirac distribution $n_{k,s}^{\rm GC}=1/(e^{\beta(\varepsilon_{k,s}-\mu)}+1)$, with $\mu$ the chemical potential and $\beta=1/T$ the inverse temperature (we adopt the standard convention $\hbar=k_B=1$).

We derive now the conditions under which the universal result Eq.~(5) can be observed in the $t_y\gg t_x$ limit. In such limit, the derivatives in Eq.~\eqref{eq:DH} can be readily calculated by first considering the expression of the single-particle spectrum~\eqref{eq:bands}:
\begin{equation}\label{eq:bandstylarge}
	\begin{split}
		\varepsilon_{\pm}(k)\stackrel{t_y\gg t_x,\nu}{\simeq}&-2t_x\cos(k+\phi)\cos\left(\frac\varphi2\right)\pm t_y\\
		&\pm\frac1{2t_y}\left[2t_x\sin(k+\phi)\sin\left(\frac\varphi2\right)+\nu\right]^2\,.
	\end{split}
\end{equation}
Taking the derivatives with respect to $\phi$ and $\nu$, one finds 
\begin{align}\label{eq:DHtylargens}
	\Delta_{\rm H}^{(t_y\gg t_x)}&=2\frac{t_x}{t_y}\frac{\sum_k \cos(k)[n_{k,+}-n_{k,-}]}{\sum_k \cos(k)[n_{k,-}+n_{k,+}]}\,. 
\end{align}
At zero temperature ($T=0$) and below half-filling ($\rho<1/2$), the upper band is necessarily empty ($n_{k,+}=0$) and one finds the universal value~(5), which we report here
\begin{equation}\label{eq:DHuniversal}
	|\Delta_{\rm H}|=\frac{2t_x}{t_y}\tan\left(\frac\varphi2\right)\,.
\end{equation}
Figure~\ref{figS7}B illustrates the progressive convergence of $\Delta_{\rm H}$ towards this universal value upon increasing $t_y$. For a fixed density $\rho$, the ground state undergoes a transition from a regime where both bands are partially filled 
to a regime with a single occupied band. 
Figure~\ref{figS7}B shows that the convergence towards the universal value is particularly fast as soon as this transition is crossed.

\begin{figure}[t!]
	\centering
	\includegraphics[width=\columnwidth]{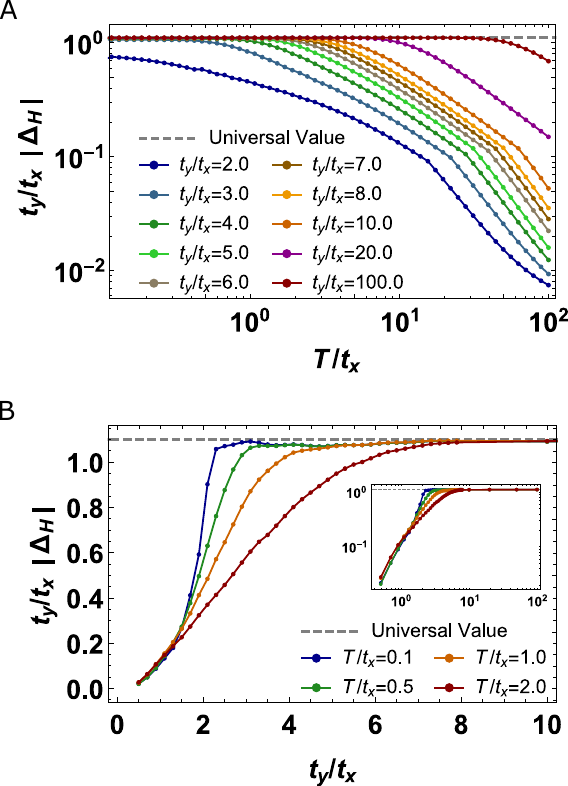}
	\caption{Robustness of the universal Hall response to finite temperatures. (\textbf{A}) Dependence of the time-averaged Hall response $\langle \Delta_{\rm H}\rangle$ on temperature $T$ for different values of $t_y>t_x$. The simulations reproduce the experimental conditions in the non-interacting limit ($U=0$), where a two-leg ladder of $L=100$ rungs, threaded by a flux $\varphi=0.32\pi$ and a confinement potential $V_x=0.01t_x$, is put out of equilibrium by  a gradient potential $E_x=0.5t_x$. The system is initially prepared in its equilibrium state in the grand-canonical ensemble, with an average number of $N=32$ fermions. (\textbf B) Same data as a function of the transverse coupling $t_y$ for different values of the temperature $T$.}
	\label{figS8}
\end{figure}

The opposite signs in front of the occupation factors $n_{k,\pm}$, appearing in the numerator of Eq.~\eqref{eq:DHtylargens}, reflect the progressive suppression of the Hall response whenever particles are added to the upper band, characterized by opposite polarization than the lower one, see also Fig.~\ref{figS7}. One thus expects that to observe the universal Hall response~\eqref{eq:DHuniversal} as long as $n_{k,+}$ remains close to zero. As a consequence, finite temperatures $T$ do not affect the universal regime as long a they remain smaller than the transverse hopping $t_y$, which also controls the separation between the upper and lower bands. The robustness of the universal Hall response~\eqref{eq:DHuniversal} with respect to temperature in non-interacting systems in the experimental conditions, including a strong drive ($E_x=0.5t_x$) and the presence of a confining potential ($V_x=0.01t_x$) is illustrated in Fig.~\ref{figS8}. In particular, these figures show that, for $T\simeq t_x$, in non-interacting systems the universal value can be observed already for $t_y\geq4t_x$. This behavior is fully consistent with the observations reported in Fig.~\ref{fig3}, where the presence of finite interactions ($U=6.56t_x$), further stabilizes the single-band metal. For this reason, the universal value~\eqref{eq:DHuniversal} is observed for lower values of $t_y\geq2t_x$. 

Notice that the definitions of the current and polarization densities, $j_x$ and $p_y$, given in Eq.~\eqref{eq:jxpy}, differ from the experimentally measured quantities, $J_x$ and $P_y$, defined in Eqs.~\eqref{eq3} and~\eqref{eq4}. When comparing theoretical calculations to experimental data, averaging over different tubes with different particle occupations, see discussion in Section~\ref{sec:no_of_atoms}, has to be taken into account.  The relation between $j_x$ and $J_x$ reads
\begin{equation}\label{eq:averaging}
	J_x=\frac L{N_{\rm tot}}\sum_{\rm tubes} j_{x,t}\,,
\end{equation}
where $L$ is the system size considered in the calculation, $N_{\rm tot}$ is the total number of particles in the system, and $j_{x,t}$ is the average current density in the tube $t$. An analogous relation holds between $p_y$ and~$P_y$. Notice that the prefactor $L/N_{\rm tot}$ disappears when taking the ratio between $P_y$ and $J_x$, giving the Hall imbalance $\Delta_{\rm H}$, defined in Eq.~(1). For this reason, the universal values~(5) and~\eqref{eq:DHuniversal} coincide. For the sake of clarity in the following theoretical analysis including interactions and confinement, we consider quantities consistent with the definition~\eqref{eq:jxpy}, and rely then on Eq.~\eqref{eq:averaging} when comparing to the experimental data.


\subsection{Stabilization of the universal regime by interactions -- Mean-field analysis}\label{sec:MF}
We now analyze the two-leg ladder in the presence of repulsive interactions of strength $U$ using a mean-field approximation (MFA). We show that repulsive interactions effectively increase the transverse coupling $t_y$ and thus stabilize the universal value of the Hall imbalance~\eqref{eq:DHuniversal}. Such approach was also recently applied to study the effects of interactions on orbital currents in these systems \cite{sHuang_2022}.

We consider $\phi=\nu=0$ and proceed with the mean-field decoupling of the interaction term in Eq.~(2)
\begin{equation}
	\begin{split}
		U\sum_{j=1}^L&n_{j,1}n_{j,0}\simeq U\sum_j\Big[n_{j,1}\av{n_{j,0}}+n_{j,0}\av{n_{j,1}}-\\
		&a^{\dagger}_{j,1}a_{j,0}\av{a^\dagger_{j,0}a_{j,1}}-a^{\dagger}_{j,0}a_{j,1}\av{a^\dagger_{j,1}a_{j,0}}\Big]\,,
	\end{split}
\end{equation}
where we have discarded the constant contributions to the total energy. If we assume equilibrium (no current flowing in the system) and that interactions do not lead to spontaneous breaking of translational invariance, averages do not depend on the lattice rung and leg labels $(j,m)$. We can thus replace the average local occupations with the density $\av{n_{j,m}}=\rho$. This substitution  leads to the standard Hartree renormalization of the chemical potential. 

Additionally, we find that interactions also lead to a renormalization of the transverse hopping $t_y$ 
\begin{align}\label{eq:typ}
	t_y&\longrightarrow t^*_y=t_y+U\,\Omega_{\varphi,U,\rho}\,,
\end{align}
with $\Omega_{\varphi,U,\rho}=\sum_j\langle a^\dagger_{j,0}a_{j,1}\rangle/L=\sum_k\langle a^\dagger_{k,0}a_{k,1}\rangle /L$. Discussing the renormalization of $t_y$ requires a self-consistent solution of the problem, which would, in any case, remain a crude approximation, as it would miss the Luttinger liquid nature of the ground state~\cite{scarrSpinlessFermionicLadders2006}. To provide additional insight, we  evaluate the function $\Omega_{\varphi,U,\rho}$ for the non-interacting problem ($U=0$), which is equivalent to first order perturbation theory in the interaction $U$: 
\begin{equation}\label{eq:omeganonint}
	\begin{split}
		\Omega_{\varphi,U,\rho}\simeq \Omega_{\varphi,\rho}=-\sum_{s=\pm}\frac{s}{2\pi} El\left[k^s_F,-\frac{4t_x^2\sin^2(\varphi/2)}{t_y^2}\right]\,,
	\end{split}
\end{equation}
where $k_F^{s}$ are the Fermi quasi-momenta of the $s=\pm$ band. $El[k,r]$ is the Elliptic function of the first kind:
\begin{equation}
	El\left[k,-\frac1a\right] \equiv \sqrt a \int_0^k \frac{dq}{\sqrt{\sin^2(q)+a}}\,,
\end{equation} 
with the properties $El\left[0,-1/a\right]=0$ and $El\left[k,0\right]=k$. For the experimentally relevant parameters, $\varphi=0.32\pi$, $t_x\approx t_y$ and $\rho<1/2$, $k_F^->k_F^+$ and thus $\Omega_{\varphi,\rho}$ is a positive quantity. Thus, at the mean-field level, repulsive interactions increase the effective transverse coupling between the chains, and stabilize the universal regime for $\Delta_{\rm H}$ (Eq.~\eqref{eq:DHuniversal}). 


\subsection{Time-dependent simulation of the Hall response with DMRG and mean-field }\label{sec:realistic time}

\begin{figure*}[t!]
    \centering
    \includegraphics[width=2\columnwidth]{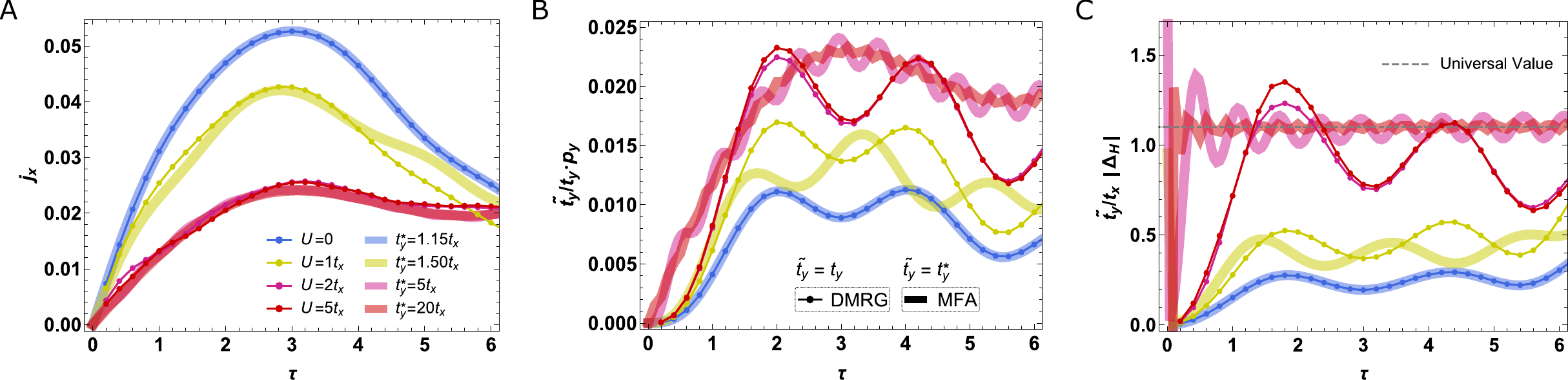}
    \caption{Real time evolution of the current (\textbf{A}), polarization density (\textbf{B}) and Hall response $\Delta_{\rm H}$ (\textbf{C}). We compare DMRG simulations for different interaction strengths $U$ (dots, where $\widetilde{t_y}=t_y$) with non-interacting systems with properly renormalized $t^*_y>t_y$ (MFA represented by thick solid lines, $\widetilde{t_y}=t^*_y$), see Eq.~\eqref{eq:typ}. Time is expressed in units of $\hbar/t_x$. We consider the experimental situation corresponding to Fig.~\ref{fig4}, namely: $t_y=1.15t_x$ and $V_x=0.01t_x$. We prepare the ground state for $N=30$ particles on $L=200$ rungs and simulate its evolution following a quench of strength $\mu_x=0.5t_x$ of a linear potential, as described in the main text.}
    \label{figS9}
\end{figure*}

The experimental conditions lie beyond the limits of our previous analysis in several ways: the strong drive $E_x =0.5 t_x$ is beyond the linear response treatment of \ref{sec:non-interacting} and the interaction strength $U$ lies beyond the perturbative regime explored in \ref{sec:MF}. Moreover, there is a parabolic confinement potential of strength $V_x=0.01t_x$, and the temperature is finite. To go beyond linear response, we calculate the time-dependent response of the system (initially prepared in the ground state of the interacting Hamiltonian Eq.~(2)) to a sudden quench of $E_x$. The interactions are taken into account exactly using zero temperature DMRG, or approximately using the mean-field approach of \ref{sec:MF}. In both cases, we use a finite parabolic confinement $V_x=0.01t_x$ to match the experimental parameters. After verifying that mean-field and DMRG results match at zero temperature, we extend our mean-field analysis to finite temperatures, inaccessible to DMRG for meaningfully large systems. Such analysis allows us to give account of the experimental observations.

\subsubsection{Real-time simulation at zero temperature: DMRG calculations}
We use the density matrix renormalization group (DMRG) algorithm~\cite{sPhysRevLett.69.2863} to obtain the ground state of the interacting two-leg ladder Eq.~(2) with additional parabolic confinement potential $V_x=0.01t_x$. At $t=0$, we quench the linear longitudinal potential from $E_x=0$ to $E_x=0.5t_x$, and simulate the real-time dynamics of the system using the time-dependent variational principle (TDVP)~\cite{sPhysRevLett.107.070601}.

\begin{figure*}[t!]
    \centering
    \includegraphics[width=2\columnwidth]{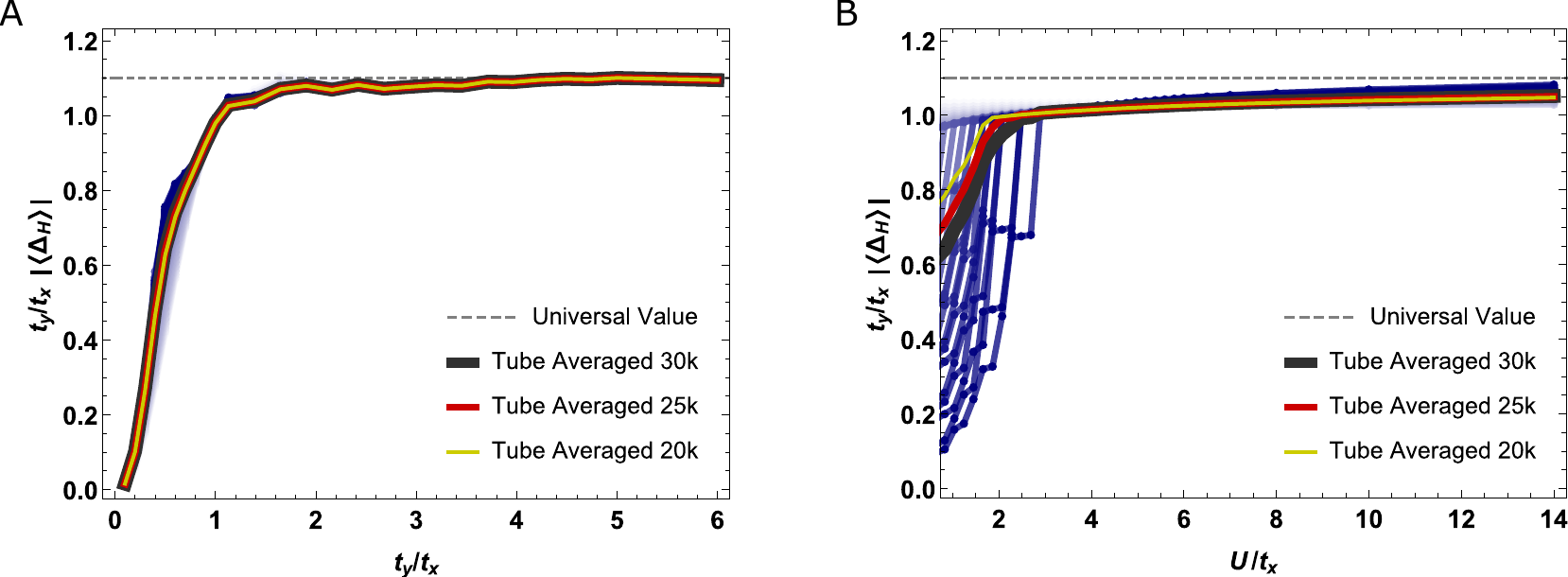}
    \caption{Time-averaged Hall response as function of $t_y/t_x$ for $U=6.56t_x$ (\textbf{A}) and as function of $U/t_x$ for $t_y=1.15t_x$ (\textbf{B}) for different number of particles. The light to dark solid lines give the Hall response for $N=1\rightarrow36$ particles on $L=200$ rungs in the same condition as Fig.~\ref{figS9}. We also show the averaged results for atom distributions corresponding to a total number of 20/25/30 thousands atoms in the system. The averages over tubes were performed for the current and the polarization separately. We then obtained the time-evolution of the Hall response $\Delta_{\rm H}(\tau)$ by taking their ratio, which was then time-averaged in the time interval $\tau\in[1,5]$, as for the experimental data presented in the main text. The simulation were performed with DMRG at zero temperature.}
    \label{figS10}
\end{figure*}

The real time evolution of the Hall response for $N=30$ spinless fermions on a two-leg ladder of $L=200$ rungs is presented in Fig.~\ref{figS9}. We have verified that all relevant observables (integrated current $L\cdot j_x$ and polarization $L\cdot p_y$) have converged with respect to $L$. We have used a bond dimension $\chi = 200$ and a time step $\delta_t =0.1/t_x $ for the time-evolution, which ensure the convergence of these observables for this system size. This choice of $N$ is representative of the average occupation of tubes in the experiment, in particular for large values of $U$, see also Fig.~\ref{figS10}.

The fact that the system is driven strongly out-of equilibrium, beyond the linear regime, results in oscillations of the current and the polarization (for weaker drives $E_x\simeq 0.01t_x$, $j_x$ and $p_y$ increase linearly with time as observed in Refs.~\cite{sPhysRevLett.122.083402,sPhysRevLett.123.086803}) . The oscillations are a combined effect of Bloch oscillations and current reversal by the confinement, and they are damped at longer times in the presence of interactions. Figure~\ref{figS9} shows that increasing interactions have a strong and opposite effect on the real time evolution of the current and of the polarization. The former is suppressed, while the latter increases (in absolute value) the larger the interaction strength $U$. Both quantities converge towards a limiting curve in the $U\gg t_x$ limit. 

Remarkably, a non-interacting theory with a properly renormalized value of the transverse coupling $t_y\rightarrow t^*_y$ quantitatively reproduces the real time evolution of the current, in agreement with the mean-field predictions of Section~\ref{sec:MF}, as illustrated in Fig.~\ref{figS9}A. The same applies for polarization but this case requires additional discussion. Relying on Eqs.~\eqref{eq:jxpy} and~\eqref{eq:bands}, one can readily show the suppression of the polarization $p_y\propto1/t_y$ in the $t_y\gg t_x$ limit, for $\rho<1/2$. As a consequence, the effective increase of $t_y\rightarrow t_y^*$ leads to an incorrect suppression of the polarization which is not observed in DMRG in the $U\gg t_x$ limit. We thus compensate this suppression by making the substitution $p_y\rightarrow p_y\cdot t^*_y/t_y $. Remarkably, this renormalized polarization qualitatively follows the exact DMRG evolution, even though it clearly displays faster oscillations. Additionally, the universal value~(5) of the Hall response $\Delta_{\rm H}$ is well reproduced by averaging over the interval $\tau\in[1,5]$ as illustrated in Fig.~\ref{fig4}.

\subsubsection{Real-time Hall response at finite-temperature}
In the experiment, the temperature is of the order of the longitudinal hopping $T\simeq t_x$. As we saw in the main text, this finite temperature has a significant influence on the Hall response in the intermediary regimes of $U$ and $t_y$, but does not affect the value of the universal Hall response at large $U$ or large $t_y/t_x$. The regime $T\simeq t_x$ and $L\simeq 200$ can hardly be accessed with DMRG for the time scales of experimental relevance. We have therefore relied on the good agreement between DMRG and a properly renormalized non-interacting theory to evaluate the effect of finite temperature in our system (Figs.~3 and~4). We stress that the results from the effective non-interacting theory should be taken with some precaution, as their agreement with exact calculations has been demonstrated exclusively for $T=0$ in the previous paragraph.

\subsubsection{Averaging over different number of tubes}
We conclude our theoretical analysis of the experimental observations by discussing the effects that the averaging over different tubes with different atom occupations may have on the observed behavior of the Hall response. In Fig.~\ref{figS10}, we show the predicted dependence of the averaged Hall imbalance $\langle \Delta_{\rm H}\rangle$ (see main text) over the transverse hopping $t_y$ and the interaction strength $U$, for different number of atoms from exact DMRG simulations at zero temperature. We also compare with the averaged over different tube distributions corresponding to a total of 20/25/30 thousand atoms, estimated as discussed in Section~\ref{sec:no_of_atoms}. The simulations clearly show that the the dependence of $\langle\Delta_{\rm H}\rangle$ over $t_y$ is much less affected than the one over the interaction strength $U$, if we consider different number of particles in the system.  The fact that stronger effects on the particle numbers are observed as function of the interactions $U$ can be explained by the fact that interactions are expected to have stronger effects, the larger the number of particles in the system.


\begin{thebibliography}{44}%
	\makeatletter
	\providecommand \@ifxundefined [1]{%
		\@ifx{#1\undefined}
	}%
	\providecommand \@ifnum [1]{%
		\ifnum #1\expandafter \@firstoftwo
		\else \expandafter \@secondoftwo
		\fi
	}%
	\providecommand \@ifx [1]{%
		\ifx #1\expandafter \@firstoftwo
		\else \expandafter \@secondoftwo
		\fi
	}%
	\providecommand \natexlab [1]{#1}%
	\providecommand \enquote  [1]{``#1''}%
	\providecommand \bibnamefont  [1]{#1}%
	\providecommand \bibfnamefont [1]{#1}%
	\providecommand \citenamefont [1]{#1}%
	\providecommand \href@noop [0]{\@secondoftwo}%
	\providecommand \href [0]{\begingroup \@sanitize@url \@href}%
	\providecommand \@href[1]{\@@startlink{#1}\@@href}%
	\providecommand \@@href[1]{\endgroup#1\@@endlink}%
	\providecommand \@sanitize@url [0]{\catcode `\\12\catcode `\$12\catcode
		`\&12\catcode `\#12\catcode `\^12\catcode `\_12\catcode `\%12\relax}%
	\providecommand \@@startlink[1]{}%
	\providecommand \@@endlink[0]{}%
	\providecommand \url  [0]{\begingroup\@sanitize@url \@url }%
	\providecommand \@url [1]{\endgroup\@href {#1}{\urlprefix }}%
	\providecommand \urlprefix  [0]{URL }%
	\providecommand \Eprint [0]{\href }%
	\providecommand \doibase [0]{https://doi.org/}%
	\providecommand \selectlanguage [0]{\@gobble}%
	\providecommand \bibinfo  [0]{\@secondoftwo}%
	\providecommand \bibfield  [0]{\@secondoftwo}%
	\providecommand \translation [1]{[#1]}%
	\providecommand \BibitemOpen [0]{}%
	\providecommand \bibitemStop [0]{}%
	\providecommand \bibitemNoStop [0]{.\EOS\space}%
	\providecommand \EOS [0]{\spacefactor3000\relax}%
	\providecommand \BibitemShut  [1]{\csname bibitem#1\endcsname}%
	\let\auto@bib@innerbib\@empty
	\bibitem [{\citenamefont {Hall}(1879)}]{Edwin}%
	\BibitemOpen
	\bibfield  {author} {\bibinfo {author} {\bibfnamefont {E.~H.}\ \bibnamefont
			{Hall}},\ }\href {http://www.jstor.org/stable/2369245} {\bibfield  {journal}
		{\bibinfo  {journal} {American Journal of Mathematics}\ }\textbf {\bibinfo
			{volume} {2}},\ \bibinfo {pages} {287} (\bibinfo {year} {1879})}\BibitemShut
	{NoStop}%
	\bibitem [{\citenamefont {Popovic}(2003)}]{popovic_hall_2003}%
	\BibitemOpen
	\bibfield  {author} {\bibinfo {author} {\bibfnamefont {R.~S.}\ \bibnamefont
			{Popovic}},\ }\href {https://doi.org/10.1201/NOE0750308557} {\emph {\bibinfo
			{title} {Hall {Effect} {Devices}}}}\ (\bibinfo  {publisher} {CRC Press, ed.
		2},\ \bibinfo {year} {2003})\BibitemShut {NoStop}%
	\bibitem [{\citenamefont {Klitzing}\ \emph {et~al.}(1980)\citenamefont
		{Klitzing}, \citenamefont {Dorda},\ and\ \citenamefont
		{Pepper}}]{PhysRevLett.45.494}%
	\BibitemOpen
	\bibfield  {author} {\bibinfo {author} {\bibfnamefont {K.~v.}\ \bibnamefont
			{Klitzing}}, \bibinfo {author} {\bibfnamefont {G.}~\bibnamefont {Dorda}},\
		and\ \bibinfo {author} {\bibfnamefont {M.}~\bibnamefont {Pepper}},\ }\href
	{https://doi.org/10.1103/PhysRevLett.45.494} {\bibfield  {journal} {\bibinfo
			{journal} {Phys. Rev. Lett.}\ }\textbf {\bibinfo {volume} {45}},\ \bibinfo
		{pages} {494} (\bibinfo {year} {1980})}\BibitemShut {NoStop}%
	\bibitem [{\citenamefont {Ong}(1991)}]{ong91_geometric_hall}%
	\BibitemOpen
	\bibfield  {author} {\bibinfo {author} {\bibfnamefont {N.~P.}\ \bibnamefont
			{Ong}},\ }\href {https://doi.org/10.1103/PhysRevB.43.193} {\bibfield
		{journal} {\bibinfo  {journal} {Phys. Rev. B}\ }\textbf {\bibinfo {volume}
			{43}},\ \bibinfo {pages} {193} (\bibinfo {year} {1991})}\BibitemShut
	{NoStop}%
	\bibitem [{\citenamefont {Tsuji}(1958)}]{tsuji58_hall_effect_cubic}%
	\BibitemOpen
	\bibfield  {author} {\bibinfo {author} {\bibfnamefont {M.}~\bibnamefont
			{Tsuji}},\ }\href {https://doi.org/10.1143/JPSJ.13.979} {\bibfield  {journal}
		{\bibinfo  {journal} {J. Phys. Soc. Jpn.}\ }\textbf {\bibinfo {volume}
			{13}},\ \bibinfo {pages} {979} (\bibinfo {year} {1958})}\BibitemShut
	{NoStop}%
	\bibitem [{\citenamefont {Xiao}\ \emph {et~al.}(2010)\citenamefont {Xiao},
		\citenamefont {Chang},\ and\ \citenamefont
		{Niu}}]{xiao2010_berry_phase_review}%
	\BibitemOpen
	\bibfield  {author} {\bibinfo {author} {\bibfnamefont {D.}~\bibnamefont
			{Xiao}}, \bibinfo {author} {\bibfnamefont {M.-C.}\ \bibnamefont {Chang}},\
		and\ \bibinfo {author} {\bibfnamefont {Q.}~\bibnamefont {Niu}},\ }\href
	{https://doi.org/10.1103/RevModPhys.82.1959} {\bibfield  {journal} {\bibinfo
			{journal} {Rev. Mod. Phys.}\ }\textbf {\bibinfo {volume} {82}},\ \bibinfo
		{pages} {1959} (\bibinfo {year} {2010})}\BibitemShut {NoStop}%
	\bibitem [{\citenamefont {Thouless}\ \emph {et~al.}(1982)\citenamefont
		{Thouless}, \citenamefont {Kohmoto}, \citenamefont {Nightingale},\ and\
		\citenamefont {den Nijs}}]{thouless1982_quantized_hall_conductance}%
	\BibitemOpen
	\bibfield  {author} {\bibinfo {author} {\bibfnamefont {D.~J.}\ \bibnamefont
			{Thouless}}, \bibinfo {author} {\bibfnamefont {M.}~\bibnamefont {Kohmoto}},
		\bibinfo {author} {\bibfnamefont {M.~P.}\ \bibnamefont {Nightingale}},\ and\
		\bibinfo {author} {\bibfnamefont {M.}~\bibnamefont {den Nijs}},\ }\href
	{https://doi.org/10.1103/PhysRevLett.49.405} {\bibfield  {journal} {\bibinfo
			{journal} {Phys. Rev. Lett.}\ }\textbf {\bibinfo {volume} {49}},\ \bibinfo
		{pages} {405} (\bibinfo {year} {1982})}\BibitemShut {NoStop}%
	\bibitem [{\citenamefont {Hasan}\ and\ \citenamefont
		{Kane}(2010)}]{RevModPhys.82.3045}%
	\BibitemOpen
	\bibfield  {author} {\bibinfo {author} {\bibfnamefont {M.~Z.}\ \bibnamefont
			{Hasan}}\ and\ \bibinfo {author} {\bibfnamefont {C.~L.}\ \bibnamefont
			{Kane}},\ }\href {https://doi.org/10.1103/RevModPhys.82.3045} {\bibfield
		{journal} {\bibinfo  {journal} {Rev. Mod. Phys.}\ }\textbf {\bibinfo {volume}
			{82}},\ \bibinfo {pages} {3045} (\bibinfo {year} {2010})}\BibitemShut
	{NoStop}%
	\bibitem [{\citenamefont {Goldman}\ \emph {et~al.}(2016)\citenamefont
		{Goldman}, \citenamefont {Budich},\ and\ \citenamefont
		{Zoller}}]{goldman2016topological}%
	\BibitemOpen
	\bibfield  {author} {\bibinfo {author} {\bibfnamefont {N.}~\bibnamefont
			{Goldman}}, \bibinfo {author} {\bibfnamefont {J.~C.}\ \bibnamefont
			{Budich}},\ and\ \bibinfo {author} {\bibfnamefont {P.}~\bibnamefont
			{Zoller}},\ }\href {https://www.nature.com/articles/nphys3803} {\bibfield
		{journal} {\bibinfo  {journal} {Nature Physics}\ }\textbf {\bibinfo {volume}
			{12}},\ \bibinfo {pages} {639} (\bibinfo {year} {2016})}\BibitemShut
	{NoStop}%
	\bibitem [{\citenamefont {Ozawa}\ \emph {et~al.}(2019)\citenamefont {Ozawa},
		\citenamefont {Price}, \citenamefont {Amo}, \citenamefont {Goldman},
		\citenamefont {Hafezi}, \citenamefont {Lu}, \citenamefont {Rechtsman},
		\citenamefont {Schuster}, \citenamefont {Simon}, \citenamefont {Zilberberg},\
		and\ \citenamefont {Carusotto}}]{RevModPhys.91.015006}%
	\BibitemOpen
	\bibfield  {author} {\bibinfo {author} {\bibfnamefont {T.}~\bibnamefont
			{Ozawa}}, \bibinfo {author} {\bibfnamefont {H.~M.}\ \bibnamefont {Price}},
		\bibinfo {author} {\bibfnamefont {A.}~\bibnamefont {Amo}}, \bibinfo {author}
		{\bibfnamefont {N.}~\bibnamefont {Goldman}}, \bibinfo {author} {\bibfnamefont
			{M.}~\bibnamefont {Hafezi}}, \bibinfo {author} {\bibfnamefont
			{L.}~\bibnamefont {Lu}}, \bibinfo {author} {\bibfnamefont {M.~C.}\
			\bibnamefont {Rechtsman}}, \bibinfo {author} {\bibfnamefont {D.}~\bibnamefont
			{Schuster}}, \bibinfo {author} {\bibfnamefont {J.}~\bibnamefont {Simon}},
		\bibinfo {author} {\bibfnamefont {O.}~\bibnamefont {Zilberberg}},\ and\
		\bibinfo {author} {\bibfnamefont {I.}~\bibnamefont {Carusotto}},\ }\href
	{https://doi.org/10.1103/RevModPhys.91.015006} {\bibfield  {journal}
		{\bibinfo  {journal} {Rev. Mod. Phys.}\ }\textbf {\bibinfo {volume} {91}},\
		\bibinfo {pages} {015006} (\bibinfo {year} {2019})}\BibitemShut {NoStop}%
	\bibitem [{\citenamefont {Tsui}\ \emph {et~al.}(1982)\citenamefont {Tsui},
		\citenamefont {Stormer},\ and\ \citenamefont {Gossard}}]{tsui_FQHE}%
	\BibitemOpen
	\bibfield  {author} {\bibinfo {author} {\bibfnamefont {D.~C.}\ \bibnamefont
			{Tsui}}, \bibinfo {author} {\bibfnamefont {H.~L.}\ \bibnamefont {Stormer}},\
		and\ \bibinfo {author} {\bibfnamefont {A.~C.}\ \bibnamefont {Gossard}},\
	}\href {https://doi.org/10.1103/PhysRevLett.48.1559} {\bibfield  {journal}
		{\bibinfo  {journal} {Phys. Rev. Lett.}\ }\textbf {\bibinfo {volume} {48}},\
		\bibinfo {pages} {1559} (\bibinfo {year} {1982})}\BibitemShut {NoStop}%
	\bibitem [{\citenamefont {Laughlin}(1983)}]{laughlin1983}%
	\BibitemOpen
	\bibfield  {author} {\bibinfo {author} {\bibfnamefont {R.~B.}\ \bibnamefont
			{Laughlin}},\ }\href {https://doi.org/10.1103/PhysRevLett.50.1395} {\bibfield
		{journal} {\bibinfo  {journal} {Phys. Rev. Lett.}\ }\textbf {\bibinfo
			{volume} {50}},\ \bibinfo {pages} {1395} (\bibinfo {year}
		{1983})}\BibitemShut {NoStop}%
	\bibitem [{\citenamefont {Yoshioka}(2002)}]{Yoshioka}%
	\BibitemOpen
	\bibfield  {author} {\bibinfo {author} {\bibfnamefont {D.}~\bibnamefont
			{Yoshioka}},\ }\href
	{https://doi.org/https://doi.org/10.1007/978-3-662-05016-3} {\emph {\bibinfo
			{title} {The Quantum Hall Effect}}},\ Vol.\ \bibinfo {volume} {133}\
	(\bibinfo  {publisher} {Springer, ed. 1},\ \bibinfo {year}
	{2002})\BibitemShut {NoStop}%
	\bibitem [{\citenamefont {Kapitulnik}\ \emph {et~al.}(2019)\citenamefont
		{Kapitulnik}, \citenamefont {Kivelson},\ and\ \citenamefont
		{Spivak}}]{RevModPhys.91.011002}%
	\BibitemOpen
	\bibfield  {author} {\bibinfo {author} {\bibfnamefont {A.}~\bibnamefont
			{Kapitulnik}}, \bibinfo {author} {\bibfnamefont {S.~A.}\ \bibnamefont
			{Kivelson}},\ and\ \bibinfo {author} {\bibfnamefont {B.}~\bibnamefont
			{Spivak}},\ }\href {https://doi.org/10.1103/RevModPhys.91.011002} {\bibfield
		{journal} {\bibinfo  {journal} {Rev. Mod. Phys.}\ }\textbf {\bibinfo {volume}
			{91}},\ \bibinfo {pages} {011002} (\bibinfo {year} {2019})}\BibitemShut
	{NoStop}%
	\bibitem [{\citenamefont {Brinkman}\ and\ \citenamefont
		{Rice}(1971)}]{PhysRevB.4.1566}%
	\BibitemOpen
	\bibfield  {author} {\bibinfo {author} {\bibfnamefont {W.~F.}\ \bibnamefont
			{Brinkman}}\ and\ \bibinfo {author} {\bibfnamefont {T.~M.}\ \bibnamefont
			{Rice}},\ }\href {https://doi.org/10.1103/PhysRevB.4.1566} {\bibfield
		{journal} {\bibinfo  {journal} {Phys. Rev. B}\ }\textbf {\bibinfo {volume}
			{4}},\ \bibinfo {pages} {1566} (\bibinfo {year} {1971})}\BibitemShut
	{NoStop}%
	\bibitem [{\citenamefont {Le\'on}\ \emph {et~al.}(2007)\citenamefont {Le\'on},
		\citenamefont {Berthod},\ and\ \citenamefont
		{Giamarchi}}]{PhysRevB.75.195123}%
	\BibitemOpen
	\bibfield  {author} {\bibinfo {author} {\bibfnamefont {G.}~\bibnamefont
			{Le\'on}}, \bibinfo {author} {\bibfnamefont {C.}~\bibnamefont {Berthod}},\
		and\ \bibinfo {author} {\bibfnamefont {T.}~\bibnamefont {Giamarchi}},\ }\href
	{https://doi.org/10.1103/PhysRevB.75.195123} {\bibfield  {journal} {\bibinfo
			{journal} {Phys. Rev. B}\ }\textbf {\bibinfo {volume} {75}},\ \bibinfo
		{pages} {195123} (\bibinfo {year} {2007})}\BibitemShut {NoStop}%
	\bibitem [{\citenamefont {Auerbach}(2018)}]{PhysRevLett.121.066601}%
	\BibitemOpen
	\bibfield  {author} {\bibinfo {author} {\bibfnamefont {A.}~\bibnamefont
			{Auerbach}},\ }\href {https://doi.org/10.1103/PhysRevLett.121.066601}
	{\bibfield  {journal} {\bibinfo  {journal} {Phys. Rev. Lett.}\ }\textbf
		{\bibinfo {volume} {121}},\ \bibinfo {pages} {066601} (\bibinfo {year}
		{2018})}\BibitemShut {NoStop}%
	\bibitem [{\citenamefont {Zotos}\ \emph {et~al.}(2000)\citenamefont {Zotos},
		\citenamefont {Naef}, \citenamefont {Long},\ and\ \citenamefont
		{Prelov\ifmmode~\check{s}\else \v{s}\fi{}ek}}]{PhysRevLett.85.377}%
	\BibitemOpen
	\bibfield  {author} {\bibinfo {author} {\bibfnamefont {X.}~\bibnamefont
			{Zotos}}, \bibinfo {author} {\bibfnamefont {F.}~\bibnamefont {Naef}},
		\bibinfo {author} {\bibfnamefont {M.}~\bibnamefont {Long}},\ and\ \bibinfo
		{author} {\bibfnamefont {P.}~\bibnamefont {Prelov\ifmmode~\check{s}\else
				\v{s}\fi{}ek}},\ }\href {https://doi.org/10.1103/PhysRevLett.85.377}
	{\bibfield  {journal} {\bibinfo  {journal} {Phys. Rev. Lett.}\ }\textbf
		{\bibinfo {volume} {85}},\ \bibinfo {pages} {377} (\bibinfo {year}
		{2000})}\BibitemShut {NoStop}%
	\bibitem [{\citenamefont {Lange}(1997)}]{PhysRevB.55.3907}%
	\BibitemOpen
	\bibfield  {author} {\bibinfo {author} {\bibfnamefont {E.}~\bibnamefont
			{Lange}},\ }\href {https://doi.org/10.1103/PhysRevB.55.3907} {\bibfield
		{journal} {\bibinfo  {journal} {Phys. Rev. B}\ }\textbf {\bibinfo {volume}
			{55}},\ \bibinfo {pages} {3907} (\bibinfo {year} {1997})}\BibitemShut
	{NoStop}%
	\bibitem [{\citenamefont {Shastry}\ \emph {et~al.}(1993)\citenamefont
		{Shastry}, \citenamefont {Shraiman},\ and\ \citenamefont
		{Singh}}]{PhysRevLett.70.2004}%
	\BibitemOpen
	\bibfield  {author} {\bibinfo {author} {\bibfnamefont {B.~S.}\ \bibnamefont
			{Shastry}}, \bibinfo {author} {\bibfnamefont {B.~I.}\ \bibnamefont
			{Shraiman}},\ and\ \bibinfo {author} {\bibfnamefont {R.~R.~P.}\ \bibnamefont
			{Singh}},\ }\href {https://doi.org/10.1103/PhysRevLett.70.2004} {\bibfield
		{journal} {\bibinfo  {journal} {Phys. Rev. Lett.}\ }\textbf {\bibinfo
			{volume} {70}},\ \bibinfo {pages} {2004} (\bibinfo {year}
		{1993})}\BibitemShut {NoStop}%
	\bibitem [{\citenamefont {Berg}\ \emph {et~al.}(2015)\citenamefont {Berg},
		\citenamefont {Huber},\ and\ \citenamefont {Lindner}}]{PhysRevB.91.024507}%
	\BibitemOpen
	\bibfield  {author} {\bibinfo {author} {\bibfnamefont {E.}~\bibnamefont
			{Berg}}, \bibinfo {author} {\bibfnamefont {S.~D.}\ \bibnamefont {Huber}},\
		and\ \bibinfo {author} {\bibfnamefont {N.~H.}\ \bibnamefont {Lindner}},\
	}\href {https://doi.org/10.1103/PhysRevB.91.024507} {\bibfield  {journal}
		{\bibinfo  {journal} {Phys. Rev. B}\ }\textbf {\bibinfo {volume} {91}},\
		\bibinfo {pages} {024507} (\bibinfo {year} {2015})}\BibitemShut {NoStop}%
	\bibitem [{\citenamefont {Hagen}\ \emph {et~al.}(1990)\citenamefont {Hagen},
		\citenamefont {Lobb}, \citenamefont {Greene}, \citenamefont {Forrester},\
		and\ \citenamefont {Kang}}]{hagen_anomalous_1990}%
	\BibitemOpen
	\bibfield  {author} {\bibinfo {author} {\bibfnamefont {S.~J.}\ \bibnamefont
			{Hagen}}, \bibinfo {author} {\bibfnamefont {C.~J.}\ \bibnamefont {Lobb}},
		\bibinfo {author} {\bibfnamefont {R.~L.}\ \bibnamefont {Greene}}, \bibinfo
		{author} {\bibfnamefont {M.~G.}\ \bibnamefont {Forrester}},\ and\ \bibinfo
		{author} {\bibfnamefont {J.~H.}\ \bibnamefont {Kang}},\ }\href
	{https://doi.org/10.1103/PhysRevB.41.11630} {\bibfield  {journal} {\bibinfo
			{journal} {Phys. Rev. B}\ }\textbf {\bibinfo {volume} {41}},\ \bibinfo
		{pages} {11630} (\bibinfo {year} {1990})}\BibitemShut {NoStop}%
	\bibitem [{\citenamefont {Badoux}\ \emph {et~al.}(2016)\citenamefont {Badoux},
		\citenamefont {Tabis}, \citenamefont {Laliberté}, \citenamefont
		{Grissonnanche}, \citenamefont {Vignolle}, \citenamefont {Vignolles},
		\citenamefont {Béard}, \citenamefont {Bonn}, \citenamefont {Hardy},
		\citenamefont {Liang}, \citenamefont {Doiron-Leyraud}, \citenamefont
		{Taillefer},\ and\ \citenamefont {Proust}}]{badoux_change_2016}%
	\BibitemOpen
	\bibfield  {author} {\bibinfo {author} {\bibfnamefont {S.}~\bibnamefont
			{Badoux}}, \bibinfo {author} {\bibfnamefont {W.}~\bibnamefont {Tabis}},
		\bibinfo {author} {\bibfnamefont {F.}~\bibnamefont {Laliberté}}, \bibinfo
		{author} {\bibfnamefont {G.}~\bibnamefont {Grissonnanche}}, \bibinfo {author}
		{\bibfnamefont {B.}~\bibnamefont {Vignolle}}, \bibinfo {author}
		{\bibfnamefont {D.}~\bibnamefont {Vignolles}}, \bibinfo {author}
		{\bibfnamefont {J.}~\bibnamefont {Béard}}, \bibinfo {author} {\bibfnamefont
			{D.~A.}\ \bibnamefont {Bonn}}, \bibinfo {author} {\bibfnamefont {W.~N.}\
			\bibnamefont {Hardy}}, \bibinfo {author} {\bibfnamefont {R.}~\bibnamefont
			{Liang}}, \bibinfo {author} {\bibfnamefont {N.}~\bibnamefont
			{Doiron-Leyraud}}, \bibinfo {author} {\bibfnamefont {L.}~\bibnamefont
			{Taillefer}},\ and\ \bibinfo {author} {\bibfnamefont {C.}~\bibnamefont
			{Proust}},\ }\href {https://www.nature.com/articles/nature16983} {\bibfield
		{journal} {\bibinfo  {journal} {Nature}\ }\textbf {\bibinfo {volume} {531}},\
		\bibinfo {pages} {210} (\bibinfo {year} {2016})}\BibitemShut {NoStop}%
	\bibitem [{\citenamefont {Smith}\ \emph {et~al.}(1994)\citenamefont {Smith},
		\citenamefont {Clinton}, \citenamefont {Tsuei},\ and\ \citenamefont
		{Lobb}}]{smith_sign_1994}%
	\BibitemOpen
	\bibfield  {author} {\bibinfo {author} {\bibfnamefont {A.~W.}\ \bibnamefont
			{Smith}}, \bibinfo {author} {\bibfnamefont {T.~W.}\ \bibnamefont {Clinton}},
		\bibinfo {author} {\bibfnamefont {C.~C.}\ \bibnamefont {Tsuei}},\ and\
		\bibinfo {author} {\bibfnamefont {C.~J.}\ \bibnamefont {Lobb}},\ }\href
	{https://doi.org/10.1103/PhysRevB.49.12927} {\bibfield  {journal} {\bibinfo
			{journal} {Phys. Rev. B}\ }\textbf {\bibinfo {volume} {49}},\ \bibinfo
		{pages} {12927} (\bibinfo {year} {1994})}\BibitemShut {NoStop}%
	\bibitem [{\citenamefont {Mih\'aly}\ \emph {et~al.}(2000)\citenamefont
		{Mih\'aly}, \citenamefont {K\'ezsm\'arki}, \citenamefont {Z\'amborszky},\
		and\ \citenamefont {Forr\'o}}]{PhysRevLett.84.2670}%
	\BibitemOpen
	\bibfield  {author} {\bibinfo {author} {\bibfnamefont {G.}~\bibnamefont
			{Mih\'aly}}, \bibinfo {author} {\bibfnamefont {I.}~\bibnamefont
			{K\'ezsm\'arki}}, \bibinfo {author} {\bibfnamefont {F.}~\bibnamefont
			{Z\'amborszky}},\ and\ \bibinfo {author} {\bibfnamefont {L.}~\bibnamefont
			{Forr\'o}},\ }\href {https://doi.org/10.1103/PhysRevLett.84.2670} {\bibfield
		{journal} {\bibinfo  {journal} {Phys. Rev. Lett.}\ }\textbf {\bibinfo
			{volume} {84}},\ \bibinfo {pages} {2670} (\bibinfo {year}
		{2000})}\BibitemShut {NoStop}%
	\bibitem [{\citenamefont {Moser}\ \emph {et~al.}(2000)\citenamefont {Moser},
		\citenamefont {Cooper}, \citenamefont {J\'erome}, \citenamefont {Alavi},
		\citenamefont {Brown},\ and\ \citenamefont
		{Bechgaard}}]{PhysRevLett.84.2674}%
	\BibitemOpen
	\bibfield  {author} {\bibinfo {author} {\bibfnamefont {J.}~\bibnamefont
			{Moser}}, \bibinfo {author} {\bibfnamefont {J.~R.}\ \bibnamefont {Cooper}},
		\bibinfo {author} {\bibfnamefont {D.}~\bibnamefont {J\'erome}}, \bibinfo
		{author} {\bibfnamefont {B.}~\bibnamefont {Alavi}}, \bibinfo {author}
		{\bibfnamefont {S.~E.}\ \bibnamefont {Brown}},\ and\ \bibinfo {author}
		{\bibfnamefont {K.}~\bibnamefont {Bechgaard}},\ }\href
	{https://doi.org/10.1103/PhysRevLett.84.2674} {\bibfield  {journal} {\bibinfo
			{journal} {Phys. Rev. Lett.}\ }\textbf {\bibinfo {volume} {84}},\ \bibinfo
		{pages} {2674} (\bibinfo {year} {2000})}\BibitemShut {NoStop}%
	\bibitem [{\citenamefont {Greschner}\ \emph {et~al.}(2019)\citenamefont
		{Greschner}, \citenamefont {Filippone},\ and\ \citenamefont
		{Giamarchi}}]{PhysRevLett.122.083402}%
	\BibitemOpen
	\bibfield  {author} {\bibinfo {author} {\bibfnamefont {S.}~\bibnamefont
			{Greschner}}, \bibinfo {author} {\bibfnamefont {M.}~\bibnamefont
			{Filippone}},\ and\ \bibinfo {author} {\bibfnamefont {T.}~\bibnamefont
			{Giamarchi}},\ }\href {https://doi.org/10.1103/PhysRevLett.122.083402}
	{\bibfield  {journal} {\bibinfo  {journal} {Phys. Rev. Lett.}\ }\textbf
		{\bibinfo {volume} {122}},\ \bibinfo {pages} {083402} (\bibinfo {year}
		{2019})}\BibitemShut {NoStop}%
	\bibitem [{\citenamefont {Dalibard}\ \emph {et~al.}(2011)\citenamefont
		{Dalibard}, \citenamefont {Gerbier}, \citenamefont
		{Juzeli\ifmmode~\bar{u}\else \={u}\fi{}nas},\ and\ \citenamefont
		{\"Ohberg}}]{RevModPhys.83.1523}%
	\BibitemOpen
	\bibfield  {author} {\bibinfo {author} {\bibfnamefont {J.}~\bibnamefont
			{Dalibard}}, \bibinfo {author} {\bibfnamefont {F.}~\bibnamefont {Gerbier}},
		\bibinfo {author} {\bibfnamefont {G.}~\bibnamefont
			{Juzeli\ifmmode~\bar{u}\else \={u}\fi{}nas}},\ and\ \bibinfo {author}
		{\bibfnamefont {P.}~\bibnamefont {\"Ohberg}},\ }\href
	{https://doi.org/10.1103/RevModPhys.83.1523} {\bibfield  {journal} {\bibinfo
			{journal} {Rev. Mod. Phys.}\ }\textbf {\bibinfo {volume} {83}},\ \bibinfo
		{pages} {1523} (\bibinfo {year} {2011})}\BibitemShut {NoStop}%
	\bibitem [{\citenamefont {Goldman}\ \emph {et~al.}(2014)\citenamefont
		{Goldman}, \citenamefont {Juzeliūnas}, \citenamefont {Öhberg},\ and\
		\citenamefont {Spielman}}]{Goldman2014}%
	\BibitemOpen
	\bibfield  {author} {\bibinfo {author} {\bibfnamefont {N.}~\bibnamefont
			{Goldman}}, \bibinfo {author} {\bibfnamefont {G.}~\bibnamefont
			{Juzeliūnas}}, \bibinfo {author} {\bibfnamefont {P.}~\bibnamefont
			{Öhberg}},\ and\ \bibinfo {author} {\bibfnamefont {I.~B.}\ \bibnamefont
			{Spielman}},\ }\href {https://doi.org/10.1088/0034-4885/77/12/126401}
	{\bibfield  {journal} {\bibinfo  {journal} {Reports on Progress in Physics}\
		}\textbf {\bibinfo {volume} {77}},\ \bibinfo {pages} {126401} (\bibinfo
		{year} {2014})}\BibitemShut {NoStop}%
	\bibitem [{\citenamefont {Cooper}\ \emph {et~al.}(2019)\citenamefont {Cooper},
		\citenamefont {Dalibard},\ and\ \citenamefont
		{Spielman}}]{RevModPhys.91.015005}%
	\BibitemOpen
	\bibfield  {author} {\bibinfo {author} {\bibfnamefont {N.~R.}\ \bibnamefont
			{Cooper}}, \bibinfo {author} {\bibfnamefont {J.}~\bibnamefont {Dalibard}},\
		and\ \bibinfo {author} {\bibfnamefont {I.~B.}\ \bibnamefont {Spielman}},\
	}\href {https://doi.org/10.1103/RevModPhys.91.015005} {\bibfield  {journal}
		{\bibinfo  {journal} {Rev. Mod. Phys.}\ }\textbf {\bibinfo {volume} {91}},\
		\bibinfo {pages} {015005} (\bibinfo {year} {2019})}\BibitemShut {NoStop}%
	\bibitem [{\citenamefont {Ozawa}\ and\ \citenamefont
		{Price}(2019)}]{Ozawa2019}%
	\BibitemOpen
	\bibfield  {author} {\bibinfo {author} {\bibfnamefont {T.}~\bibnamefont
			{Ozawa}}\ and\ \bibinfo {author} {\bibfnamefont {H.~M.}\ \bibnamefont
			{Price}},\ }\href {https://doi.org/10.1038/s42254-019-0045-3} {\bibfield
		{journal} {\bibinfo  {journal} {Nature Reviews Physics}\ }\textbf {\bibinfo
			{volume} {1}},\ \bibinfo {pages} {349} (\bibinfo {year} {2019})}\BibitemShut
	{NoStop}%
	\bibitem [{\citenamefont {Mancini}\ \emph {et~al.}(2015)\citenamefont
		{Mancini}, \citenamefont {Pagano}, \citenamefont {Cappellini}, \citenamefont
		{Livi}, \citenamefont {Rider}, \citenamefont {Catani}, \citenamefont {Sias},
		\citenamefont {Zoller}, \citenamefont {Inguscio}, \citenamefont {Dalmonte},\
		and\ \citenamefont {Fallani}}]{Mancini1510}%
	\BibitemOpen
	\bibfield  {author} {\bibinfo {author} {\bibfnamefont {M.}~\bibnamefont
			{Mancini}}, \bibinfo {author} {\bibfnamefont {G.}~\bibnamefont {Pagano}},
		\bibinfo {author} {\bibfnamefont {G.}~\bibnamefont {Cappellini}}, \bibinfo
		{author} {\bibfnamefont {L.}~\bibnamefont {Livi}}, \bibinfo {author}
		{\bibfnamefont {M.}~\bibnamefont {Rider}}, \bibinfo {author} {\bibfnamefont
			{J.}~\bibnamefont {Catani}}, \bibinfo {author} {\bibfnamefont
			{C.}~\bibnamefont {Sias}}, \bibinfo {author} {\bibfnamefont {P.}~\bibnamefont
			{Zoller}}, \bibinfo {author} {\bibfnamefont {M.}~\bibnamefont {Inguscio}},
		\bibinfo {author} {\bibfnamefont {M.}~\bibnamefont {Dalmonte}},\ and\
		\bibinfo {author} {\bibfnamefont {L.}~\bibnamefont {Fallani}},\ }\href
	{https://doi.org/10.1126/science.aaa8736} {\bibfield  {journal} {\bibinfo
			{journal} {Science}\ }\textbf {\bibinfo {volume} {349}},\ \bibinfo {pages}
		{1510} (\bibinfo {year} {2015})}\BibitemShut {NoStop}%
	\bibitem [{\citenamefont {Aidelsburger}\ \emph {et~al.}(2015)\citenamefont
		{Aidelsburger}, \citenamefont {Lohse}, \citenamefont {Schweizer},
		\citenamefont {Atala}, \citenamefont {Barreiro}, \citenamefont {Nascimbène},
		\citenamefont {Cooper}, \citenamefont {Bloch},\ and\ \citenamefont
		{Goldman}}]{Aidelsburger2015}%
	\BibitemOpen
	\bibfield  {author} {\bibinfo {author} {\bibfnamefont {M.}~\bibnamefont
			{Aidelsburger}}, \bibinfo {author} {\bibfnamefont {M.}~\bibnamefont {Lohse}},
		\bibinfo {author} {\bibfnamefont {C.}~\bibnamefont {Schweizer}}, \bibinfo
		{author} {\bibfnamefont {M.}~\bibnamefont {Atala}}, \bibinfo {author}
		{\bibfnamefont {J.~T.}\ \bibnamefont {Barreiro}}, \bibinfo {author}
		{\bibfnamefont {S.}~\bibnamefont {Nascimbène}}, \bibinfo {author}
		{\bibfnamefont {N.~R.}\ \bibnamefont {Cooper}}, \bibinfo {author}
		{\bibfnamefont {I.}~\bibnamefont {Bloch}},\ and\ \bibinfo {author}
		{\bibfnamefont {N.}~\bibnamefont {Goldman}},\ }\href
	{https://doi.org/10.1038/nphys3171} {\bibfield  {journal} {\bibinfo
			{journal} {Nature Physics}\ }\textbf {\bibinfo {volume} {11}},\ \bibinfo
		{pages} {162} (\bibinfo {year} {2015})}\BibitemShut {NoStop}%
	\bibitem [{\citenamefont {Genkina}\ \emph {et~al.}(2019)\citenamefont
		{Genkina}, \citenamefont {Aycock}, \citenamefont {Lu}, \citenamefont {Lu},
		\citenamefont {Pineiro},\ and\ \citenamefont
		{Spielman}}]{genkina2019imaging}%
	\BibitemOpen
	\bibfield  {author} {\bibinfo {author} {\bibfnamefont {D.}~\bibnamefont
			{Genkina}}, \bibinfo {author} {\bibfnamefont {L.~M.}\ \bibnamefont {Aycock}},
		\bibinfo {author} {\bibfnamefont {H.-I.}\ \bibnamefont {Lu}}, \bibinfo
		{author} {\bibfnamefont {M.}~\bibnamefont {Lu}}, \bibinfo {author}
		{\bibfnamefont {A.~M.}\ \bibnamefont {Pineiro}},\ and\ \bibinfo {author}
		{\bibfnamefont {I.}~\bibnamefont {Spielman}},\ }\href
	{https://iopscience.iop.org/article/10.1088/1367-2630/ab165b} {\bibfield
		{journal} {\bibinfo  {journal} {New journal of physics}\ }\textbf {\bibinfo
			{volume} {21}},\ \bibinfo {pages} {053021} (\bibinfo {year}
		{2019})}\BibitemShut {NoStop}%
	\bibitem [{\citenamefont {Chalopin}\ \emph {et~al.}(2020)\citenamefont
		{Chalopin}, \citenamefont {Satoor}, \citenamefont {Evrard}, \citenamefont
		{Makhalov}, \citenamefont {Dalibard}, \citenamefont {Lopes},\ and\
		\citenamefont {Nascimbene}}]{chalopin_probing_2020}%
	\BibitemOpen
	\bibfield  {author} {\bibinfo {author} {\bibfnamefont {T.}~\bibnamefont
			{Chalopin}}, \bibinfo {author} {\bibfnamefont {T.}~\bibnamefont {Satoor}},
		\bibinfo {author} {\bibfnamefont {A.}~\bibnamefont {Evrard}}, \bibinfo
		{author} {\bibfnamefont {V.}~\bibnamefont {Makhalov}}, \bibinfo {author}
		{\bibfnamefont {J.}~\bibnamefont {Dalibard}}, \bibinfo {author}
		{\bibfnamefont {R.}~\bibnamefont {Lopes}},\ and\ \bibinfo {author}
		{\bibfnamefont {S.}~\bibnamefont {Nascimbene}},\ }\href
	{https://doi.org/10.1038/s41567-020-0942-5} {\bibfield  {journal} {\bibinfo
			{journal} {Nature Physics}\ }\textbf {\bibinfo {volume} {16}},\ \bibinfo
		{pages} {1017} (\bibinfo {year} {2020})}\BibitemShut {NoStop}%
	\bibitem [{\citenamefont {Tai}\ \emph {et~al.}(2017)\citenamefont {Tai},
		\citenamefont {Lukin}, \citenamefont {Rispoli}, \citenamefont {Schittko},
		\citenamefont {Menke}, \citenamefont {Dan}, \citenamefont {Preiss},
		\citenamefont {Grusdt}, \citenamefont {Kaufman},\ and\ \citenamefont
		{Greiner}}]{Greiner2017}%
	\BibitemOpen
	\bibfield  {author} {\bibinfo {author} {\bibfnamefont {M.~E.}\ \bibnamefont
			{Tai}}, \bibinfo {author} {\bibfnamefont {A.}~\bibnamefont {Lukin}}, \bibinfo
		{author} {\bibfnamefont {M.}~\bibnamefont {Rispoli}}, \bibinfo {author}
		{\bibfnamefont {R.}~\bibnamefont {Schittko}}, \bibinfo {author}
		{\bibfnamefont {T.}~\bibnamefont {Menke}}, \bibinfo {author} {\bibfnamefont
			{B.}~\bibnamefont {Dan}}, \bibinfo {author} {\bibfnamefont {P.~M.}\
			\bibnamefont {Preiss}}, \bibinfo {author} {\bibfnamefont {F.}~\bibnamefont
			{Grusdt}}, \bibinfo {author} {\bibfnamefont {A.~M.}\ \bibnamefont
			{Kaufman}},\ and\ \bibinfo {author} {\bibfnamefont {M.}~\bibnamefont
			{Greiner}},\ }\href {https://doi.org/10.1038/nature22811} {\bibfield
		{journal} {\bibinfo  {journal} {Nature}\ }\textbf {\bibinfo {volume} {546}},\
		\bibinfo {pages} {519} (\bibinfo {year} {2017})}\BibitemShut {NoStop}%
	\bibitem [{\citenamefont {Léonard}\ \emph {et~al.}(2023)\citenamefont
		{Léonard}, \citenamefont {Kim}, \citenamefont {Kwan}, \citenamefont
		{Segura}, \citenamefont {Grusdt}, \citenamefont {Repellin}, \citenamefont
		{Goldman},\ and\ \citenamefont {Greiner}}]{Greiner2023}%
	\BibitemOpen
	\bibfield  {author} {\bibinfo {author} {\bibfnamefont {J.}~\bibnamefont
			{Léonard}}, \bibinfo {author} {\bibfnamefont {S.}~\bibnamefont {Kim}},
		\bibinfo {author} {\bibfnamefont {J.}~\bibnamefont {Kwan}}, \bibinfo {author}
		{\bibfnamefont {P.}~\bibnamefont {Segura}}, \bibinfo {author} {\bibfnamefont
			{F.}~\bibnamefont {Grusdt}}, \bibinfo {author} {\bibfnamefont
			{C.}~\bibnamefont {Repellin}}, \bibinfo {author} {\bibfnamefont
			{N.}~\bibnamefont {Goldman}},\ and\ \bibinfo {author} {\bibfnamefont
			{M.}~\bibnamefont {Greiner}},\ }\bibfield  {journal} {\bibinfo  {journal}
		{Nature}\ }\href {https://doi.org/10.1038/s41586-023-06122-4}
	{10.1038/s41586-023-06122-4} (\bibinfo {year} {2023})\BibitemShut {NoStop}%
	\bibitem [{SS()}]{SS}%
	\BibitemOpen
	\href@noop {} {\bibinfo  {journal} {See supplementary materials}\
	}\BibitemShut {NoStop}%
	\bibitem [{\citenamefont {Taie}\ \emph {et~al.}(2010)\citenamefont {Taie},
		\citenamefont {Takasu}, \citenamefont {Sugawa}, \citenamefont {Yamazaki},
		\citenamefont {Tsujimoto}, \citenamefont {Murakami},\ and\ \citenamefont
		{Takahashi}}]{PhysRevLett.105.190401}%
	\BibitemOpen
	\bibfield  {journal} {  }\bibfield  {author} {\bibinfo {author} {\bibfnamefont
			{S.}~\bibnamefont {Taie}}, \bibinfo {author} {\bibfnamefont {Y.}~\bibnamefont
			{Takasu}}, \bibinfo {author} {\bibfnamefont {S.}~\bibnamefont {Sugawa}},
		\bibinfo {author} {\bibfnamefont {R.}~\bibnamefont {Yamazaki}}, \bibinfo
		{author} {\bibfnamefont {T.}~\bibnamefont {Tsujimoto}}, \bibinfo {author}
		{\bibfnamefont {R.}~\bibnamefont {Murakami}},\ and\ \bibinfo {author}
		{\bibfnamefont {Y.}~\bibnamefont {Takahashi}},\ }\href
	{https://doi.org/10.1103/PhysRevLett.105.190401} {\bibfield  {journal}
		{\bibinfo  {journal} {Phys. Rev. Lett.}\ }\textbf {\bibinfo {volume} {105}},\
		\bibinfo {pages} {190401} (\bibinfo {year} {2010})}\BibitemShut {NoStop}%
	\bibitem [{\citenamefont {White}(1992)}]{PhysRevLett.69.2863}%
	\BibitemOpen
	\bibfield  {author} {\bibinfo {author} {\bibfnamefont {S.~R.}\ \bibnamefont
			{White}},\ }\href {https://doi.org/10.1103/PhysRevLett.69.2863} {\bibfield
		{journal} {\bibinfo  {journal} {Phys. Rev. Lett.}\ }\textbf {\bibinfo
			{volume} {69}},\ \bibinfo {pages} {2863} (\bibinfo {year}
		{1992})}\BibitemShut {NoStop}%
	\bibitem [{\citenamefont {Buser}\ \emph {et~al.}(2021)\citenamefont {Buser},
		\citenamefont {Greschner}, \citenamefont {Schollw\"ock},\ and\ \citenamefont
		{Giamarchi}}]{buser2021}%
	\BibitemOpen
	\bibfield  {author} {\bibinfo {author} {\bibfnamefont {M.}~\bibnamefont
			{Buser}}, \bibinfo {author} {\bibfnamefont {S.}~\bibnamefont {Greschner}},
		\bibinfo {author} {\bibfnamefont {U.}~\bibnamefont {Schollw\"ock}},\ and\
		\bibinfo {author} {\bibfnamefont {T.}~\bibnamefont {Giamarchi}},\ }\href
	{https://doi.org/10.1103/PhysRevLett.126.030501} {\bibfield  {journal}
		{\bibinfo  {journal} {Phys. Rev. Lett.}\ }\textbf {\bibinfo {volume} {126}},\
		\bibinfo {pages} {030501} (\bibinfo {year} {2021})}\BibitemShut {NoStop}%
	\bibitem [{\citenamefont {Hauschild}\ and\ \citenamefont
		{Pollmann}(2018)}]{tenpy1}%
	\BibitemOpen
	\bibfield  {author} {\bibinfo {author} {\bibfnamefont {J.}~\bibnamefont
			{Hauschild}}\ and\ \bibinfo {author} {\bibfnamefont {F.}~\bibnamefont
			{Pollmann}},\ }\href {https://doi.org/10.21468/SciPostPhysLectNotes.5}
	{\bibfield  {journal} {\bibinfo  {journal} {SciPost Phys. Lect. Notes}\
		}\textbf {\bibinfo {volume} {2018}},\ \bibinfo {pages} {5} (\bibinfo {year}
		{2018})}\BibitemShut {NoStop}%
	\bibitem [{ten()}]{tenpy2}%
	\BibitemOpen
	\href@noop {} {\bibfield  {journal} {\bibinfo  {journal} {TeNPy code;}\
	}}\bibinfo {note} {\url{https://github.com/tenpy/tenpy}}\BibitemShut
	{NoStop}%
	\bibitem [{\citenamefont {Zhou}\ \emph {et~al.}(2023)\citenamefont {Zhou},
		\citenamefont {Cappellini}, \citenamefont {Tusi}, \citenamefont {Franchi},
		\citenamefont {Parravicini}, \citenamefont {Repellin}, \citenamefont
		{Greschner}, \citenamefont {Inguscio}, \citenamefont {Giamarchi},
		\citenamefont {Filippone}, \citenamefont {Catani},\ and\ \citenamefont
		{Fallani}}]{Zenodo}%
	\BibitemOpen
	\bibfield  {author} {\bibinfo {author} {\bibfnamefont {T.-W.}\ \bibnamefont
			{Zhou}}, \bibinfo {author} {\bibfnamefont {G.}~\bibnamefont {Cappellini}},
		\bibinfo {author} {\bibfnamefont {D.}~\bibnamefont {Tusi}}, \bibinfo {author}
		{\bibfnamefont {L.}~\bibnamefont {Franchi}}, \bibinfo {author} {\bibfnamefont
			{J.}~\bibnamefont {Parravicini}}, \bibinfo {author} {\bibfnamefont
			{C.}~\bibnamefont {Repellin}}, \bibinfo {author} {\bibfnamefont
			{S.}~\bibnamefont {Greschner}}, \bibinfo {author} {\bibfnamefont
			{M.}~\bibnamefont {Inguscio}}, \bibinfo {author} {\bibfnamefont
			{T.}~\bibnamefont {Giamarchi}}, \bibinfo {author} {\bibfnamefont
			{M.}~\bibnamefont {Filippone}}, \bibinfo {author} {\bibfnamefont
			{J.}~\bibnamefont {Catani}},\ and\ \bibinfo {author} {\bibfnamefont
			{L.}~\bibnamefont {Fallani}},\ }\href
	{https://doi.org/10.5281/zenodo.7485725} {10.5281/zenodo.7485725} (\bibinfo
	{year} {2023})\BibitemShut {NoStop}%
\end{thebibliography}

\begin{thebibliography}{13}%
	\makeatletter
	\providecommand \@ifxundefined [1]{%
		\@ifx{#1\undefined}
	}%
	\providecommand \@ifnum [1]{%
		\ifnum #1\expandafter \@firstoftwo
		\else \expandafter \@secondoftwo
		\fi
	}%
	\providecommand \@ifx [1]{%
		\ifx #1\expandafter \@firstoftwo
		\else \expandafter \@secondoftwo
		\fi
	}%
	\providecommand \natexlab [1]{#1}%
	\providecommand \enquote  [1]{``#1''}%
	\providecommand \bibnamefont  [1]{#1}%
	\providecommand \bibfnamefont [1]{#1}%
	\providecommand \citenamefont [1]{#1}%
	\providecommand \href@noop [0]{\@secondoftwo}%
	\providecommand \href [0]{\begingroup \@sanitize@url \@href}%
	\providecommand \@href[1]{\@@startlink{#1}\@@href}%
	\providecommand \@@href[1]{\endgroup#1\@@endlink}%
	\providecommand \@sanitize@url [0]{\catcode `\\12\catcode `\$12\catcode
		`\&12\catcode `\#12\catcode `\^12\catcode `\_12\catcode `\%12\relax}%
	\providecommand \@@startlink[1]{}%
	\providecommand \@@endlink[0]{}%
	\providecommand \url  [0]{\begingroup\@sanitize@url \@url }%
	\providecommand \@url [1]{\endgroup\@href {#1}{\urlprefix }}%
	\providecommand \urlprefix  [0]{URL }%
	\providecommand \Eprint [0]{\href }%
	\providecommand \doibase [0]{https://doi.org/}%
	\providecommand \selectlanguage [0]{\@gobble}%
	\providecommand \bibinfo  [0]{\@secondoftwo}%
	\providecommand \bibfield  [0]{\@secondoftwo}%
	\providecommand \translation [1]{[#1]}%
	\providecommand \BibitemOpen [0]{}%
	\providecommand \bibitemStop [0]{}%
	\providecommand \bibitemNoStop [0]{.\EOS\space}%
	\providecommand \EOS [0]{\spacefactor3000\relax}%
	\providecommand \BibitemShut  [1]{\csname bibitem#1\endcsname}%
	\let\auto@bib@innerbib\@empty
	\bibitem [{\citenamefont {Taie}\ \emph {et~al.}(2010)\citenamefont {Taie},
		\citenamefont {Takasu}, \citenamefont {Sugawa}, \citenamefont {Yamazaki},
		\citenamefont {Tsujimoto}, \citenamefont {Murakami},\ and\ \citenamefont
		{Takahashi}}]{sPhysRevLett.105.190401}%
	\BibitemOpen
	\bibfield  {author} {\bibinfo {author} {\bibfnamefont {S.}~\bibnamefont
			{Taie}}, \bibinfo {author} {\bibfnamefont {Y.}~\bibnamefont {Takasu}},
		\bibinfo {author} {\bibfnamefont {S.}~\bibnamefont {Sugawa}}, \bibinfo
		{author} {\bibfnamefont {R.}~\bibnamefont {Yamazaki}}, \bibinfo {author}
		{\bibfnamefont {T.}~\bibnamefont {Tsujimoto}}, \bibinfo {author}
		{\bibfnamefont {R.}~\bibnamefont {Murakami}},\ and\ \bibinfo {author}
		{\bibfnamefont {Y.}~\bibnamefont {Takahashi}},\ }\href
	{https://doi.org/10.1103/PhysRevLett.105.190401} {\bibfield  {journal}
		{\bibinfo  {journal} {Phys. Rev. Lett.}\ }\textbf {\bibinfo {volume} {105}},\
		\bibinfo {pages} {190401} (\bibinfo {year} {2010})}\BibitemShut {NoStop}%
	\bibitem [{\citenamefont {Shin}\ \emph {et~al.}(2004)\citenamefont {Shin},
		\citenamefont {Saba}, \citenamefont {Pasquini}, \citenamefont {Ketterle},
		\citenamefont {Pritchard},\ and\ \citenamefont
		{Leanhardt}}]{sPhysRevLett.92.050405}%
	\BibitemOpen
	\bibfield  {author} {\bibinfo {author} {\bibfnamefont {Y.}~\bibnamefont
			{Shin}}, \bibinfo {author} {\bibfnamefont {M.}~\bibnamefont {Saba}}, \bibinfo
		{author} {\bibfnamefont {T.~A.}\ \bibnamefont {Pasquini}}, \bibinfo {author}
		{\bibfnamefont {W.}~\bibnamefont {Ketterle}}, \bibinfo {author}
		{\bibfnamefont {D.~E.}\ \bibnamefont {Pritchard}},\ and\ \bibinfo {author}
		{\bibfnamefont {A.~E.}\ \bibnamefont {Leanhardt}},\ }\href
	{https://doi.org/10.1103/PhysRevLett.92.050405} {\bibfield  {journal}
		{\bibinfo  {journal} {Phys. Rev. Lett.}\ }\textbf {\bibinfo {volume} {92}},\
		\bibinfo {pages} {050405} (\bibinfo {year} {2004})}\BibitemShut {NoStop}%
	\bibitem [{\citenamefont {Carr}\ \emph {et~al.}(2004)\citenamefont {Carr},
		\citenamefont {Shlyapnikov},\ and\ \citenamefont
		{Castin}}]{sPhysRevLett.92.150404}%
	\BibitemOpen
	\bibfield  {author} {\bibinfo {author} {\bibfnamefont {L.~D.}\ \bibnamefont
			{Carr}}, \bibinfo {author} {\bibfnamefont {G.~V.}\ \bibnamefont
			{Shlyapnikov}},\ and\ \bibinfo {author} {\bibfnamefont {Y.}~\bibnamefont
			{Castin}},\ }\href {https://doi.org/10.1103/PhysRevLett.92.150404} {\bibfield
		{journal} {\bibinfo  {journal} {Phys. Rev. Lett.}\ }\textbf {\bibinfo
			{volume} {92}},\ \bibinfo {pages} {150404} (\bibinfo {year}
		{2004})}\BibitemShut {NoStop}%
	\bibitem [{\citenamefont {Ho}\ and\ \citenamefont {Zhou}(2009)}]{sHo2009}%
	\BibitemOpen
	\bibfield  {author} {\bibinfo {author} {\bibfnamefont {T.-L.}\ \bibnamefont
			{Ho}}\ and\ \bibinfo {author} {\bibfnamefont {Q.}~\bibnamefont {Zhou}},\
	}\href {https://doi.org/10.1073/pnas.0809862105} {\bibfield  {journal}
		{\bibinfo  {journal} {Proceedings of the National Academy of Sciences}\
		}\textbf {\bibinfo {volume} {106}},\ \bibinfo {pages} {6916} (\bibinfo {year}
		{2009})}\BibitemShut {NoStop}%
	\bibitem [{\citenamefont {Taie}\ \emph {et~al.}(2012)\citenamefont {Taie},
		\citenamefont {Yamazaki}, \citenamefont {Sugawa},\ and\ \citenamefont
		{Takahashi}}]{sTaie2012}%
	\BibitemOpen
	\bibfield  {author} {\bibinfo {author} {\bibfnamefont {S.}~\bibnamefont
			{Taie}}, \bibinfo {author} {\bibfnamefont {R.}~\bibnamefont {Yamazaki}},
		\bibinfo {author} {\bibfnamefont {S.}~\bibnamefont {Sugawa}},\ and\ \bibinfo
		{author} {\bibfnamefont {Y.}~\bibnamefont {Takahashi}},\ }\href
	{https://doi.org/10.1038/nphys2430} {\bibfield  {journal} {\bibinfo
			{journal} {Nature Physics}\ }\textbf {\bibinfo {volume} {8}},\ \bibinfo
		{pages} {825} (\bibinfo {year} {2012})}\BibitemShut {NoStop}%
	\bibitem [{\citenamefont {Zotos}\ \emph {et~al.}(2000)\citenamefont {Zotos},
		\citenamefont {Naef}, \citenamefont {Long},\ and\ \citenamefont
		{Prelov\ifmmode~\check{s}\else \v{s}\fi{}ek}}]{sPhysRevLett.85.377}%
	\BibitemOpen
	\bibfield  {author} {\bibinfo {author} {\bibfnamefont {X.}~\bibnamefont
			{Zotos}}, \bibinfo {author} {\bibfnamefont {F.}~\bibnamefont {Naef}},
		\bibinfo {author} {\bibfnamefont {M.}~\bibnamefont {Long}},\ and\ \bibinfo
		{author} {\bibfnamefont {P.}~\bibnamefont {Prelov\ifmmode~\check{s}\else
				\v{s}\fi{}ek}},\ }\href {https://doi.org/10.1103/PhysRevLett.85.377}
	{\bibfield  {journal} {\bibinfo  {journal} {Phys. Rev. Lett.}\ }\textbf
		{\bibinfo {volume} {85}},\ \bibinfo {pages} {377} (\bibinfo {year}
		{2000})}\BibitemShut {NoStop}%
	\bibitem [{\citenamefont {Prelov\ifmmode~\check{s}\else \v{s}\fi{}ek}\ \emph
		{et~al.}(1999)\citenamefont {Prelov\ifmmode~\check{s}\else \v{s}\fi{}ek},
		\citenamefont {Long}, \citenamefont {Marke\ifmmode~\check{z}\else
			\v{z}\fi{}},\ and\ \citenamefont {Zotos}}]{sPhysRevLett.83.2785}%
	\BibitemOpen
	\bibfield  {author} {\bibinfo {author} {\bibfnamefont {P.}~\bibnamefont
			{Prelov\ifmmode~\check{s}\else \v{s}\fi{}ek}}, \bibinfo {author}
		{\bibfnamefont {M.}~\bibnamefont {Long}}, \bibinfo {author} {\bibfnamefont
			{T.}~\bibnamefont {Marke\ifmmode~\check{z}\else \v{z}\fi{}}},\ and\ \bibinfo
		{author} {\bibfnamefont {X.}~\bibnamefont {Zotos}},\ }\href
	{https://doi.org/10.1103/PhysRevLett.83.2785} {\bibfield  {journal} {\bibinfo
			{journal} {Phys. Rev. Lett.}\ }\textbf {\bibinfo {volume} {83}},\ \bibinfo
		{pages} {2785} (\bibinfo {year} {1999})}\BibitemShut {NoStop}%
	\bibitem [{\citenamefont {Greschner}\ \emph {et~al.}(2019)\citenamefont
		{Greschner}, \citenamefont {Filippone},\ and\ \citenamefont
		{Giamarchi}}]{sPhysRevLett.122.083402}%
	\BibitemOpen
	\bibfield  {author} {\bibinfo {author} {\bibfnamefont {S.}~\bibnamefont
			{Greschner}}, \bibinfo {author} {\bibfnamefont {M.}~\bibnamefont
			{Filippone}},\ and\ \bibinfo {author} {\bibfnamefont {T.}~\bibnamefont
			{Giamarchi}},\ }\href {https://doi.org/10.1103/PhysRevLett.122.083402}
	{\bibfield  {journal} {\bibinfo  {journal} {Phys. Rev. Lett.}\ }\textbf
		{\bibinfo {volume} {122}},\ \bibinfo {pages} {083402} (\bibinfo {year}
		{2019})}\BibitemShut {NoStop}%
	\bibitem [{\citenamefont {Filippone}\ \emph {et~al.}(2019)\citenamefont
		{Filippone}, \citenamefont {Bardyn}, \citenamefont {Greschner},\ and\
		\citenamefont {Giamarchi}}]{sPhysRevLett.123.086803}%
	\BibitemOpen
	\bibfield  {author} {\bibinfo {author} {\bibfnamefont {M.}~\bibnamefont
			{Filippone}}, \bibinfo {author} {\bibfnamefont {C.-E.}\ \bibnamefont
			{Bardyn}}, \bibinfo {author} {\bibfnamefont {S.}~\bibnamefont {Greschner}},\
		and\ \bibinfo {author} {\bibfnamefont {T.}~\bibnamefont {Giamarchi}},\ }\href
	{https://doi.org/10.1103/PhysRevLett.123.086803} {\bibfield  {journal}
		{\bibinfo  {journal} {Phys. Rev. Lett.}\ }\textbf {\bibinfo {volume} {123}},\
		\bibinfo {pages} {086803} (\bibinfo {year} {2019})}\BibitemShut {NoStop}%
	\bibitem [{\citenamefont {Huang}\ \emph {et~al.}(2022)\citenamefont {Huang},
		\citenamefont {Tezuka},\ and\ \citenamefont {Cazalilla}}]{sHuang_2022}%
	\BibitemOpen
	\bibfield  {author} {\bibinfo {author} {\bibfnamefont {C.-H.}\ \bibnamefont
			{Huang}}, \bibinfo {author} {\bibfnamefont {M.}~\bibnamefont {Tezuka}},\ and\
		\bibinfo {author} {\bibfnamefont {M.~A.}\ \bibnamefont {Cazalilla}},\ }\href
	{https://doi.org/10.1088/1367-2630/ac5a87} {\bibfield  {journal} {\bibinfo
			{journal} {New Journal of Physics}\ }\textbf {\bibinfo {volume} {24}},\
		\bibinfo {pages} {033043} (\bibinfo {year} {2022})}\BibitemShut {NoStop}%
	\bibitem [{\citenamefont {Carr}\ \emph {et~al.}(2006)\citenamefont {Carr},
		\citenamefont {Narozhny},\ and\ \citenamefont
		{Nersesyan}}]{scarrSpinlessFermionicLadders2006}%
	\BibitemOpen
	\bibfield  {author} {\bibinfo {author} {\bibfnamefont {S.~T.}\ \bibnamefont
			{Carr}}, \bibinfo {author} {\bibfnamefont {B.~N.}\ \bibnamefont {Narozhny}},\
		and\ \bibinfo {author} {\bibfnamefont {A.~A.}\ \bibnamefont {Nersesyan}},\
	}\href {https://doi.org/10.1103/PhysRevB.73.195114} {\bibfield  {journal}
		{\bibinfo  {journal} {Phys. Rev. B}\ }\textbf {\bibinfo {volume} {73}},\
		\bibinfo {pages} {195114} (\bibinfo {year} {2006})}\BibitemShut {NoStop}%
	\bibitem [{\citenamefont {White}(1992)}]{sPhysRevLett.69.2863}%
	\BibitemOpen
	\bibfield  {author} {\bibinfo {author} {\bibfnamefont {S.~R.}\ \bibnamefont
			{White}},\ }\href {https://doi.org/10.1103/PhysRevLett.69.2863} {\bibfield
		{journal} {\bibinfo  {journal} {Phys. Rev. Lett.}\ }\textbf {\bibinfo
			{volume} {69}},\ \bibinfo {pages} {2863} (\bibinfo {year}
		{1992})}\BibitemShut {NoStop}%
	\bibitem [{\citenamefont {Haegeman}\ \emph {et~al.}(2011)\citenamefont
		{Haegeman}, \citenamefont {Cirac}, \citenamefont {Osborne}, \citenamefont
		{Pi\ifmmode~\check{z}\else \v{z}\fi{}orn}, \citenamefont {Verschelde},\ and\
		\citenamefont {Verstraete}}]{sPhysRevLett.107.070601}%
	\BibitemOpen
	\bibfield  {author} {\bibinfo {author} {\bibfnamefont {J.}~\bibnamefont
			{Haegeman}}, \bibinfo {author} {\bibfnamefont {J.~I.}\ \bibnamefont {Cirac}},
		\bibinfo {author} {\bibfnamefont {T.~J.}\ \bibnamefont {Osborne}}, \bibinfo
		{author} {\bibfnamefont {I.}~\bibnamefont {Pi\ifmmode~\check{z}\else
				\v{z}\fi{}orn}}, \bibinfo {author} {\bibfnamefont {H.}~\bibnamefont
			{Verschelde}},\ and\ \bibinfo {author} {\bibfnamefont {F.}~\bibnamefont
			{Verstraete}},\ }\href {https://doi.org/10.1103/PhysRevLett.107.070601}
	{\bibfield  {journal} {\bibinfo  {journal} {Phys. Rev. Lett.}\ }\textbf
		{\bibinfo {volume} {107}},\ \bibinfo {pages} {070601} (\bibinfo {year}
		{2011})}\BibitemShut {NoStop}%
\end{thebibliography}
\end{document}